\documentclass[usenatbib]{basi}
\usepackage[T1]{fontenc}
\usepackage[british]{babel}
\usepackage[varg]{txfonts}

\usepackage{rotating}
\usepackage{dcolumn}

\usepackage{natbib,twoopt}
\usepackage[breaklinks=true]{hyperref} 
\bibpunct{(}{)}{;}{a}{}{,} 
\newcommandtwoopt{\citeads}[3][][]{\href{http://adsabs.harvard.edu/abs/#3}%
{\citealp[#1][#2]{#3}}}
\newcommandtwoopt{\citepads}[3][][]{\href{http://adsabs.harvard.edu/abs/#3}%
{\citep[#1][#2]{#3}}}
\newcommandtwoopt{\citetads}[3][][]{\href{http://adsabs.harvard.edu/abs/#3}%
{\citet[#1][#2]{#3}}}
\newcommandtwoopt{\citeyearads}[3][][]%
{\href{http://adsabs.harvard.edu/abs/#3}{\citeyear[#1][#2]{#3}}}

\makeatletter\NAT@longnamesfalse\makeatother

\begin{document}

\title[High-resolution spectroscopy and high-density monitoring in X-rays of Novae]{X-ray spectroscopy and monitoring of Novae\thanks{Using data obtained with XMM-Newton (an ESA science mission with instruments and contributions directly funded by ESA Member States and NASA), Chandra, and Swift}}
\author[J.~U.~Ness]%
       {J.~U.~Ness\thanks{email: \texttt{juness@sciops.esa.int}}\\
       XMM-Newton Science Operations Centre, ESA, PO Box 78, 28691 Villanueva de la Ca\~nada, Madrid, Spain\\
}
\pubyear{2012}
\volume{00}
\pagerange{\pageref{firstpage}--\pageref{lastpage}}

\date{Received --- ; accepted ---}

\maketitle
\label{firstpage}

\begin{abstract}

The 21st century X-ray observatories XMM-Newton, Chandra, and Swift
gave us completely new insights into the X-ray behaviour of
nova outbursts. These new-generation X-ray observatories provide
particularly high spectral resolution and high density in
monitoring campaigns, simultaneously in X-rays and UV/optical.
The entire evolution of several nova outbursts has been observed
with the Swift XRT and UVOT instruments, allowing studies of the 
gradual shift of the peak of the SED from UV to X-rays, time
scales to the onset and duration of the X-ray brightest supersoft
source (SSS) phase, and pre- and post-SSS X-ray emission.
In addition, XMM-Newton and Chandra observations can efficiently
be scheduled, allowing deeper studies of strategically chosen
evolutionary stages. Before Swift joined in 2005, Chandra and
XMM-Newton blind shots in search of SSS emission unavoidably
led to some underexposed observations taken before and/or after the SSS
phase. More systematic Swift studies reduced this number while
increasing the number of novae.
Pre- and post-SSS spectra at low and high spectral resolution
were successfully modelled with collisional plasma models.
Pre-SSS emission arises in shocks and post-SSS emission
in radiatively cooling thin ejecta. In contrast, the grating
spectra taken during the SSS phase are a lot more complex than
expected and have not yet been successfully modeled.
Available hot WD radiation transport models give
only approximate reproduction of the observations, and make
some critical assumptions that are only valid in isolated
WDs. More grating spectra which would be important to search
for systematic trends between SSS spectra and system parameters.\\

\noindent
Summary of well-established discoveries with Swift, XMM-Newton, and Chandra:\\
$\bullet$ About 50\% of novae display faint X-ray emission
  before the start of the SSS phase\\
$\bullet$ The start of the SSS phase is not a smooth process. High-amplitude
variations during the early SSS phase were seen that disappear
close to the time when the optical plateau phase begins.\\
$\bullet$ The end of the SSS phase is in most cases a smooth process.\\
$\bullet$ The SSS grating spectra contain continuum spectra that
 roughly resemble a blackbody shape\\
$\bullet$ The SSS X-ray grating spectra of systems with known high inclination angles contain emission lines on top of the continuum\\
$\bullet$ The SSS X-ray spectra of systems with unknown or low inclination
angles contain deep absorption lines from the interstellar medium and
local, high-ionisation absorption lines that are blue shifted.

\end{abstract}


\section{Introduction}\label{s:intro}

 Classical Novae (CNe) occur in cataclysmic binary systems (CVs) in which
the close companion star provides H-rich nuclear burning fuel that is
transferred to the surface of the white dwarf (WD) primary via accretion.
Once enough hydrogen-rich
material is accumulated, a thermonuclear runaway triggers the initial
explosion. The radiative energy output exceeds the Eddington limit and
drives an optically thick wind. The way we observe a nova from Earth
depends on the properties of the ejected material.
Absorption and scattering change the colour of the light that
originates from the central nuclear burning processes. During radiative
transport through the optically thick part of the ejecta, the peak of the
broad-band spectrum (or spectral energy distribution, SED), is
shifted to successively lower energy. Observed optical light arises
after radiation transport through a larger column than X-ray
emission. The opacity evolution of the
ejecta may or may not be a homogeneous process, but the bolometric
luminosity should be dictated by the energy budget of central nuclear
burning that is believed to be constant until all hydrogen burning material
has been consumed or ejected \citep{gallstar76}. With the continuously shifting SED,
observations of nova evolution require multiwavelength monitoring,
including the UV and X-ray bands. The last evolutionary phase before a
nova turns off is the supersoft source (SSS) phase, during which
atmospheric X-ray emission from the central regions of the ejecta,
close to the surface of the WD, can be observed. The start of the
SSS phase depends on the opacity evolution of the ejecta and can not
be predicted accurately. Before and after the SSS phase, novae are
X-ray faint, and scheduling X-ray observations yields a risk of
non-detections if observed too early or too late. Recent dense
X-ray/UV monitoring observations obtained with the Swift satellite
give important insights into the long-term opacity evolution of the
ejecta, also facilitating more accurate scheduling of deeper X-ray
observations, employing the grating spectrometers on board
XMM-Newton and Chandra. X-ray grating spectra yield higher
{\it spectral} resolution, allowing determinations of physical
properties such as the energetics of nuclear burning, mass of the
underlying WD, and chemical composition. Followup X-ray observations
are also of longer continuous durations, allowing studies of
short-term ($10^2-10^4$\,s) variability and associated spectral changes.\\

 In this review, I summarise observing strategies to catch the
SSS phase, at first via educated guesses based on theoretical
expectations and observational evidence from optical
({Sect.~\ref{s:preswift}), and later assisted with Swift monitoring
observations (Sect.~\ref{s:swassist}). I only touch on some selected 
nova observations to exemplify the key points of observing
strategies and data analysis. More general conclusions from the
entirety of all observations are presented in Sect.~\ref{s:concl}.

\section{A brief history of past X-ray observations of novae}\label{s:past}

 Before the advent of Chandra and XMM-Newton, instruments of
moderate sensitivity and spectral resolution but large fields
of view were available. In surveys of the Large Magellanic
Cloud (LMC) with the Einstein Observatory, a few mysterious
luminous sources with extremely soft spectra were discovered
\citep*{cal_discovery} which were later combined in a new
class of Super Soft Sources (SSS) after more had been discovered
with ROSAT \citep*{sssclass,sssclass1}. The underlying systems
were identified as accreting WDs with ongoing
nuclear burning on their surface \citep*{kahab}. Since novae
occur in such systems, the concept of constant bolometric
luminosity throughout the evolution with a continuous shift
of the peak of the SED towards higher energies, predicts that
they emit SSS X-ray spectra during their late, X-ray bright,
phase \citep{gallstar76}. This was confirmed by, e.g., EXOSAT and
ROSAT observations \citep*{oegelman87}. Monitoring observations
over time intervals of 2-3 years resulted in rough light
curves that could be divided into a rise phase, plateau phase,
and decline phase \citep*{krautt96}.\\

 The available X-ray detectors of the time allowed no precise
measurements of photon energies, leaving uncertainties of
several 10-100\,eV. Nevertheless, the term 'X-ray spectrum' is
occasionally used for 30-100 channel distributions of
photon energies. Obviously, no detailed spectral modelling is
possible with such data, and the first spectral characterisations
consisted of mere blackbody fits yielding effective temperatures
of several 10\,eV (a few $10^5$\,K) and bolometric luminosities
near or even beyond the Eddington limit (for electrons) of a
1-M$_\odot$ WD. Such high values of luminosity, several months to
years after the initial outburst, imply an unphysically high total
mass loss. In light of the incorrect assumption of a blackbody model
(thermal equilibrium), the corresponding best-fit parameters
have no physical meaning.
In an attempt to derive more physically meaningful results,
atmosphere models were applied. For example,
\cite*{balm98} determined a lower luminosity from LTE
atmosphere model fits. However, including
further more physics of non-LTE models, yield different
luminosities again, e.g., \cite*{hartheis97}. When calling in
mind the small number of observed energy channels, one may
wonder whether atmosphere models with their high degree of
complexity are sufficiently constrained by the data.\\
{\em The dilemma of the low-resolution spectra is that good fits
to the data can be achieved with either well-constrained but
unphysical models or with more physical but unconstrained models.
None of these approaches can lead us to truly informative results.}

Substantially higher spectral resolution is undoubtedly needed,
and the Chandra and XMM-Newton missions with their grating
spectrometers (see Sect.~\ref{s:ins:grat}) are a great leap
forward. This article concentrates on the Swift and XMM-Newton/Chandra
light curves and spectra. For more information on early X-ray
observations of novae, I refer to reviews by
\cite*{krautt02,krautt08,st08,bodeevansbook}.\\

\section{Current-day X-ray observing facilities}
\label{s:ins}

 In the following subsections, the instrumentation of XMM-Newton,
Chandra, and Swift is described in brevity.

\subsection{XMM-Newton and Chandra}
\label{s:ins:grat}

Chandra and XMM-Newton carry CCD spectrometers with similar energy
resolution of previous missions but with higher sensitivity and
spatial resolution. In addition, much higher spectral resolution
can be achieved with the new gratings. The dispersed photons are
recorded by X-ray sensitive detectors whose spatial resolution
determines the spectral resolution of the extracted spectrum.
XMM-Newton carries three X-ray mirrors with one
directing all light on a single pn CCD detector array
\citep{epic_pn} while the light from the other two mirrors
are shared between Metal Oxide Semi-conductor (MOS) detectors 
\citep{epic_mos} and Reflection Grating Spectrometers (RGS;
\citealt{rgs}). The wavelength range of RGS1 and RGS2 is
6-38\,\AA, and the wavelength resolution is $\sim 0.06$\,\AA.\\
 The Chandra mission carries two grating spectrometers,
a Low- and a High- Energy Transmission Grating (LETGS and HETGS),
covering the wavelength ranges 1-175\,\AA\ and 1-35\,\AA,
with wavelength resolutions of 0.06 and 0.01\,\AA, respectively.
As recording devices, a microchannel high-resolution image
(HRI) detector and chip arrays of the Advanced CCD Imaging
Spectrometer (ACIS) are available and can be combined depending
on scientific demands. The LETGS/HRC-S combination covers a wavelength
range from 1-175\,\AA\ at 0.06\,\AA\ resolution, and the HETGS/ACIS-S
combination covers 1-38\,\AA\ at 0.02\,\AA\ resolution. The LETGS/ACIS-S
combination (1-38\,\AA\ at 0.06\,\AA\ resolution) is sometimes used
for higher effective areas at 20-38\,\AA\ at the expense of spectral
resolution compared to HETGS/ACIS-S and wavelength coverage compared
to LETGS/HRC-S. The ACIS detector allows order sorting, thus the
removal of higher-dispersion order photons overlapping with first-order
photons, taking advantage of the energy resolution of the ACIS. The HRC has not
sufficient energy resolution for order sorting but suffers no
pileup. Order sorting is also possible with the XMM-Newton/RGS.

\subsection{Swift $\gamma$/UV/X-ray Observatory}
\label{s:swift}

The primary purpose of Swift is to detect $\gamma$-ray bursts (GRB)
and carry out rapid follow up observations in the X-ray and UV bands.
The satellite was thus designed with extremely short slew times between
distant targets in the sky. The required instrumentation needs to be
sensitive from $\gamma$ rays via X-rays, down to UV and
optical. A GRB detected by the Burst Alert Telescope (BAT) can then
be followed up with the X-Ray Telescope (XRT; \citealt{xrt}) and
finally with Ultra-Violet/Optical Telescope (UVOT; \citealt{uvot}).
The XRT and UVOT resemble the Metal Oxide Semi-conductor (MOS) and
the Optical Monitor (OM) on board XMM-Newton, respectively.
The XRT allows a more precise
determination of the position of a GRB and rough spectral
parameters such as flux and neutral hydrogen column density,
$N_{\rm H}$. Deeper observations with more sensitive observatories
such as Chandra or XMM-Newton, but also by Suzaku, can be
planned, and adequate exposure times can be calculated.\\

Owing to it's low-Earth orbit of $\sim 96$\,minutes, Swift can
take no long, continuous observations. The maximum visibility
for the majority of targets is $\sim 2$\,ks (depending on
declination). Deeper Swift observations
therefore consist of multiple short snapshots. Photometric
studies are only possible for variations on time scales of days
or on very short time scales of seconds to minutes.\\

 Swift detects about 90 GRBs a year and spends some $\sim 20-30$\%
of the available observing time for followups (Kim Page, priv. comm).
More than half of the Swift time budget can be filled
with observations of other targets. The low-Earth orbit allows
only short snapshots of a few hundred seconds every 96 minutes,
each yielding an X-ray and UV/optical flux plus some spectral
information. While the resulting X-ray spectra can not compete
against those obtained from long, uninterrupted Chandra or
XMM-Newton observations, large series of Swift snapshots over
extended periods of time are of high value to monitor the
X-ray and UV evolution of variable sources such as supernovae,
novae, and X-ray binaries, in unprecedented detail.

\section{X-ray Observations of Novae}
\label{s:obs}

A large number of Swift, XMM-Newton, and Chandra observations
of novae have been taken, not all of which can be discussed
in this review. Early Swift observations focused on single
detections of older novae, while a prolific monitoring
program was developed after the initial monitoring of the
prestigious recurrent nova RS\,Oph was extremely successful.
Some examples are described in Sect.~\ref{s:swmonitoring}.
Before these monitoring observations were done, XMM-Newton and Chandra
had to dig in the mist to catch the important SSS phase
with the X-ray gratings, which is described in
Sect.~\ref{s:preswift}. The time by which the SSS phase
should start, and thus the best time for deep X-ray grating
observations, is not well constrained by observations in
other wavelength bands. For example, an indicator for the
presence of SSS emission could be observations of
high-ionisation optical lines \citep{schwarz2011}.
Swift-assisted grating observations
are much more efficient, which is described in
Sect.~\ref{s:swassist}. I touch on individual novae,
describing the respective observing strategies and some
results from data analysis. Beyond the examples given here,
there are a lot more, and for further reading, I recommend
\cite{schwarz2011}. I focus on the spectral analysis with only
brief mention of additional timing analysis.

\subsection{Swift observations}
\label{s:swmonitoring}

The first Swift observations of novae addressed the question whether
or not any significant post-outburst emission was still present from
previous novae. For example, V1494\,Aql (see Sect.~\ref{s:preswift})
was observed
three times, more than six years after outburst, and was not detected.
Meanwhile, V4743\,Sgr that was observed four years after outburst,
yielded a clear detection. The old nova V723\,Cas from 1995 had never
been observed in X-rays until Swift offered a first observation at
lower price than Chandra or XMM-Newton. The continuous rise of
high-ionisation emission lines in optical spectra had suggested the
presence of a hot photo ionising source which could be directly
visible in X-rays \citep{schwarz2011}. Indeed, a clear SSS spectrum
was found, 14 years after outburst \citep{ness_v723}. The further
evolution was followed with more Swift observations and later also
with two XMM-Newton observations. V723\,Cas is now the record holder
with the latest X-ray detection on day 6018, 16.5 years after outburst.\\

A first summary of X-ray observations with Swift in comparison to
previous programs is shown in the left panel of Fig.~\ref{f:swcmp},
taken from \cite{swnovae}. At that time, multiple X-ray observations of
the same nova were rather the exception than the rule, and the
sample included a large range of speed classes for different novae.
An updated description of the Swift archive of nova observations
was presented by \cite{schwarz2011}, focusing on comparisons of
the large monitoring programs. In the right panel of Fig.~\ref{f:swcmp},
colour codes are used to mark different phases based on spectral
shapes and brightness, as described in the caption. The arrangement
in vertical direction is an indicator for the speed class, using the
full width at half maximum (FWHM) line width of optical lines (values
given in right and left).
With narrower FWHM (thus slower speed class), the SSS phase occurs
later in the evolution.\\


\begin{figure}[htbp]
\resizebox{\hsize}{!}{\includegraphics{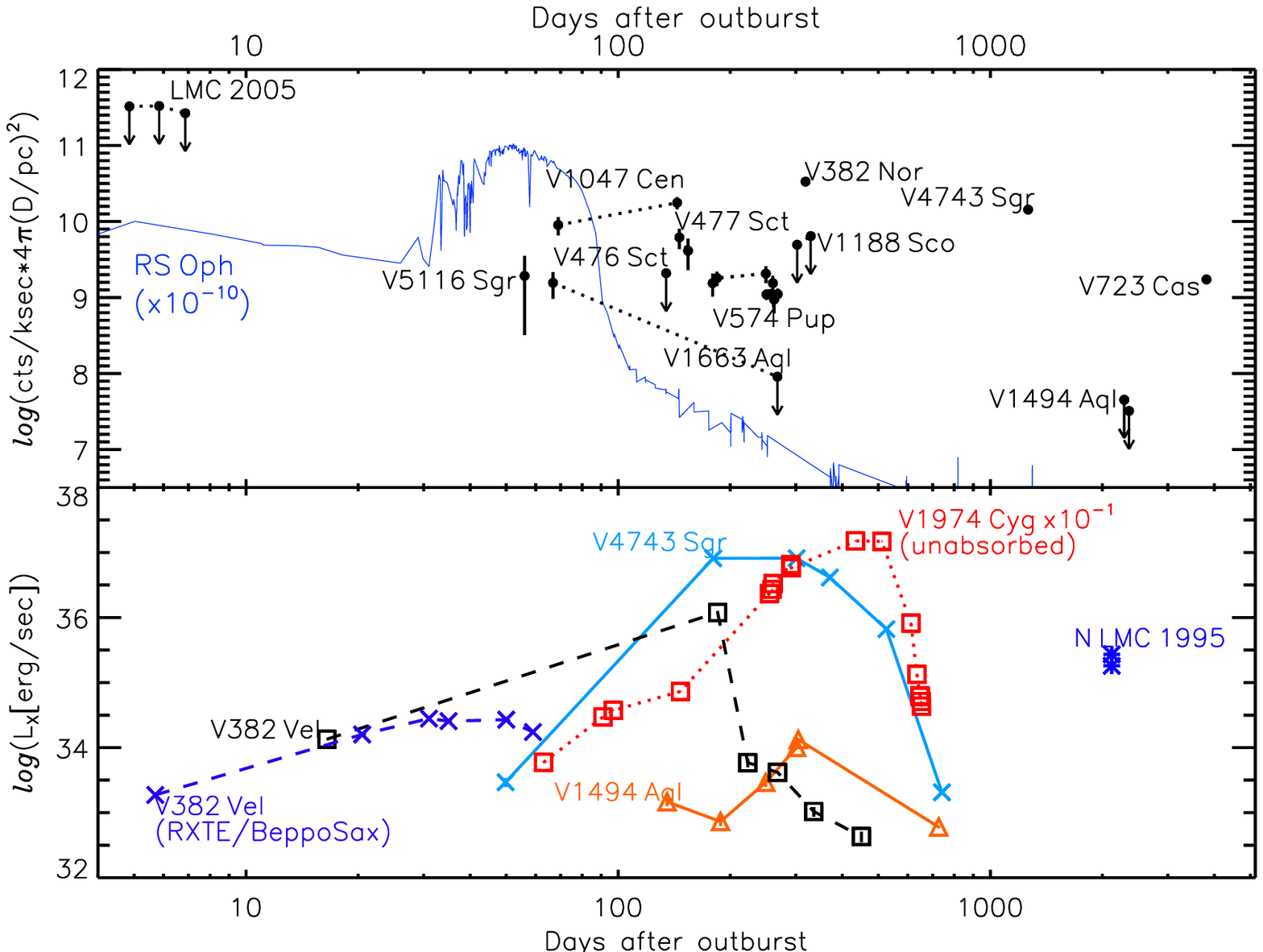}\includegraphics[angle=90,scale=0.90]{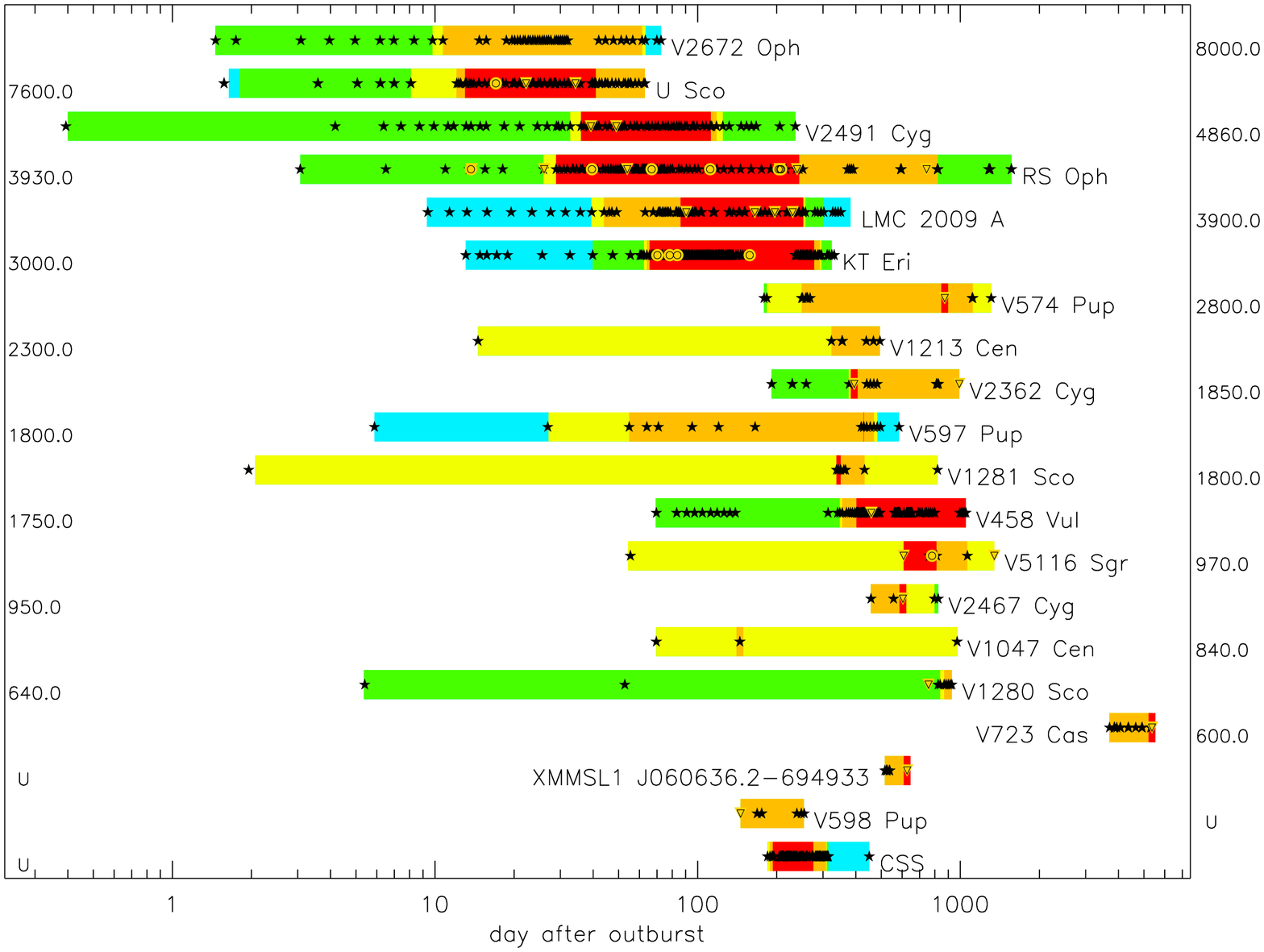}}
\caption{Overview of Swift monitoring observations of novae.
{\bf Left} from \cite{swnovae}: X-ray fluxes as function of time after
respective outburst. The top panel shows the Swift data available in
2007 and the bottom one the same for data from other missions.
The RS\,Oph Swift light curve (that was not included in \citealt{swnovae})
is added in purple for comparison.
{\bf Right} from \cite{schwarz2011}: Overview of results from Swift
monitoring campaigns of novae in which SSS emission was observed. The
colours are: light blue: non-detections, green: detections but no
clear or hard spectrum, yellow: possible SSS emission, orange weak SSS emission,
red: strong SSS emission. The plot symbols indicate times when
observations were taken with Swift (stars), XMM-Newton (triangles),
and Chandra (circles).
The novae are arranged by increasing optical emission line FWHM with
values shown either left or right of the source.  "U" is used
for novae with unknown FWHM velocities. For larger versions, see
original articles.
\label{f:swcmp}}
\end{figure}

\begin{figure}[!ht]
\resizebox{\hsize}{!}{\includegraphics{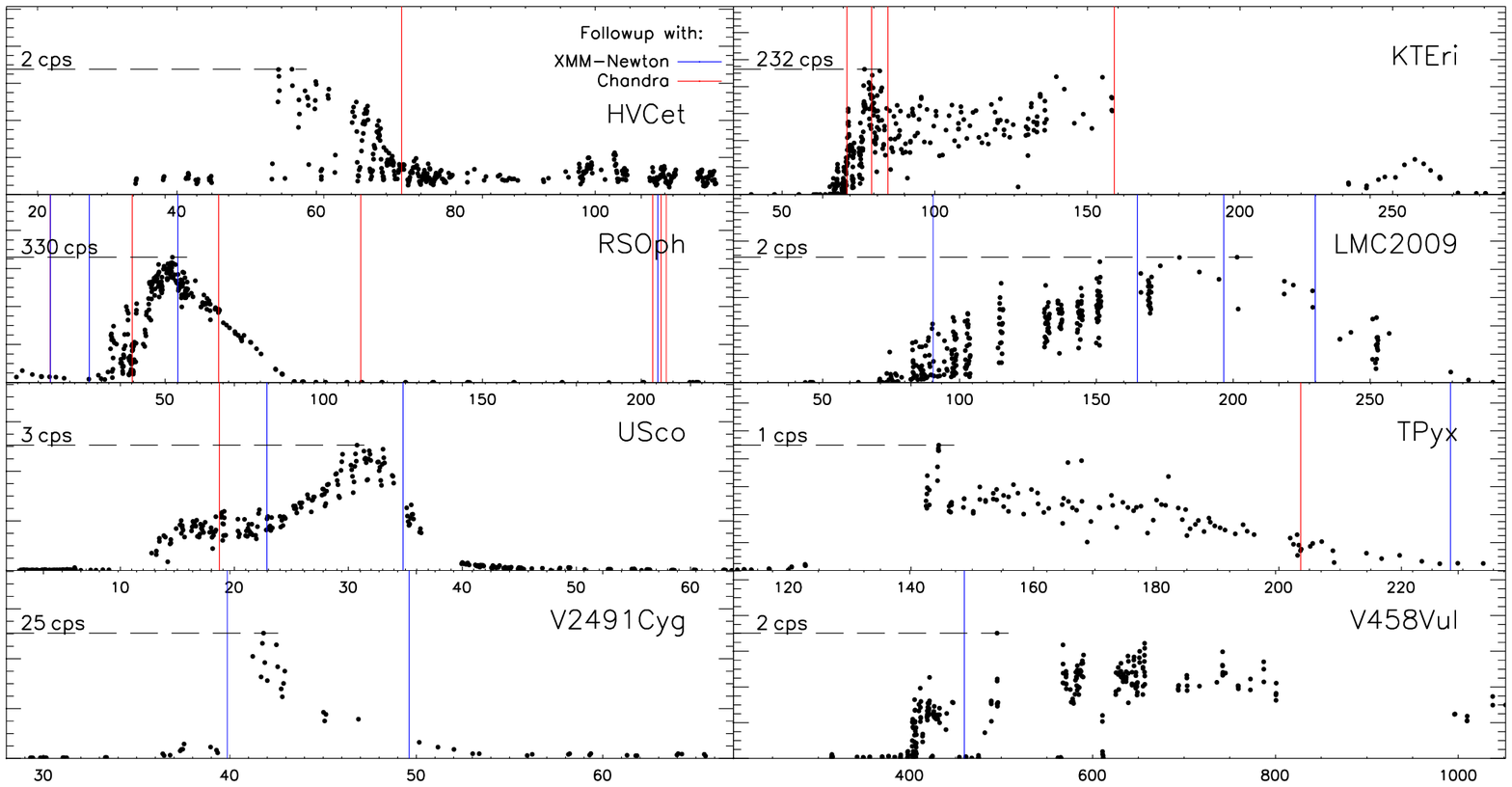}}
\caption{\label{lc}Swift X-ray monitoring observations of novae
result in unprecedented long-term light curves.
The best 8 examples are shown with the name of each nova given
in the top right of each panel and the maximum observed count
rate (cps=counts per second) marked in the left above a horizontal
dashed line. The horizontal time axes are different for each novae.
A large range of time scales and brightness levels have been
encountered. The coloured vertical lines indicate the times
of XMM-Newton (blue) and Chandra (red) observations.
Optimal times and exposure time for deeper followup observations
can be determined from the Swift monitoring, making XMM-Newton and
Chandra observations a lot more valueable.
}
\end{figure}

The long-term Swift X-ray (and partly simultaneous UV) light curves
are presented, e.g., in \cite{schwarz2011}, eight of these are shown
in Fig.~\ref{lc}, where XMM-Newton and Chandra observations are
marked by blue and orange lines, respectively (see below). Each plot
has a different y-scale, and to give an impression of the respective
brightness levels, the maximum count rates are marked
with a horizontal dashed line with values in units
counts per second (cps) given in the left. Two orders
of magnitude in count rate have been encountered.

The first dense Swift X-ray monitoring programme of a nova
was carried out following the 6th outburst of the recurrent
symbiotic nova RS\,Oph in February 2006 (\citealt{page08,osborne11},
second panel in left column in Fig.~\ref{lc}).
The X-ray brightness evolution was followed until long after the
nova had turned off. The early shock phase, the SSS phase, and
the decline phases were all extremely well covered with brightness
and spectral information. An illustration of the available Swift
data is shown in the top plot in Fig.~\ref{f:rsoph_swmap}.
The early shock phase can be seen in the spectral brightness
map above 1\,keV, while the later SSS phase covers the range
between 0.3-0.9\,keV in energy and 30-90\,days after outburst
in time. The early shock phase was analysed
by \cite{bode06}, and the evolution of shock velocity was
derived, using the plasma temperature that was derived from
collisional plasma models to the Swift spectra. Later,
\cite{vaytet2011} used hydrodynamic models yielding
much higher shock velocities, indicating that the gas
temperature from standard models is not the ideal way to
assess the shock velocity. In addition to high shock
velocities, a rather low ejecta mass was found. The models
were only applied to the low-resolution Swift spectra,
excluding the SSS component in the later observations.
While the standard models by \cite{bode06} also reproduce
the high-resolution grating spectra \citep{rsophshock}, this
has not yet been tested with the hydro model used by \cite{vaytet2011}.\\

\begin{figure}[!ht]
\begin{centering}
\resizebox{8.5cm}{!}{\includegraphics{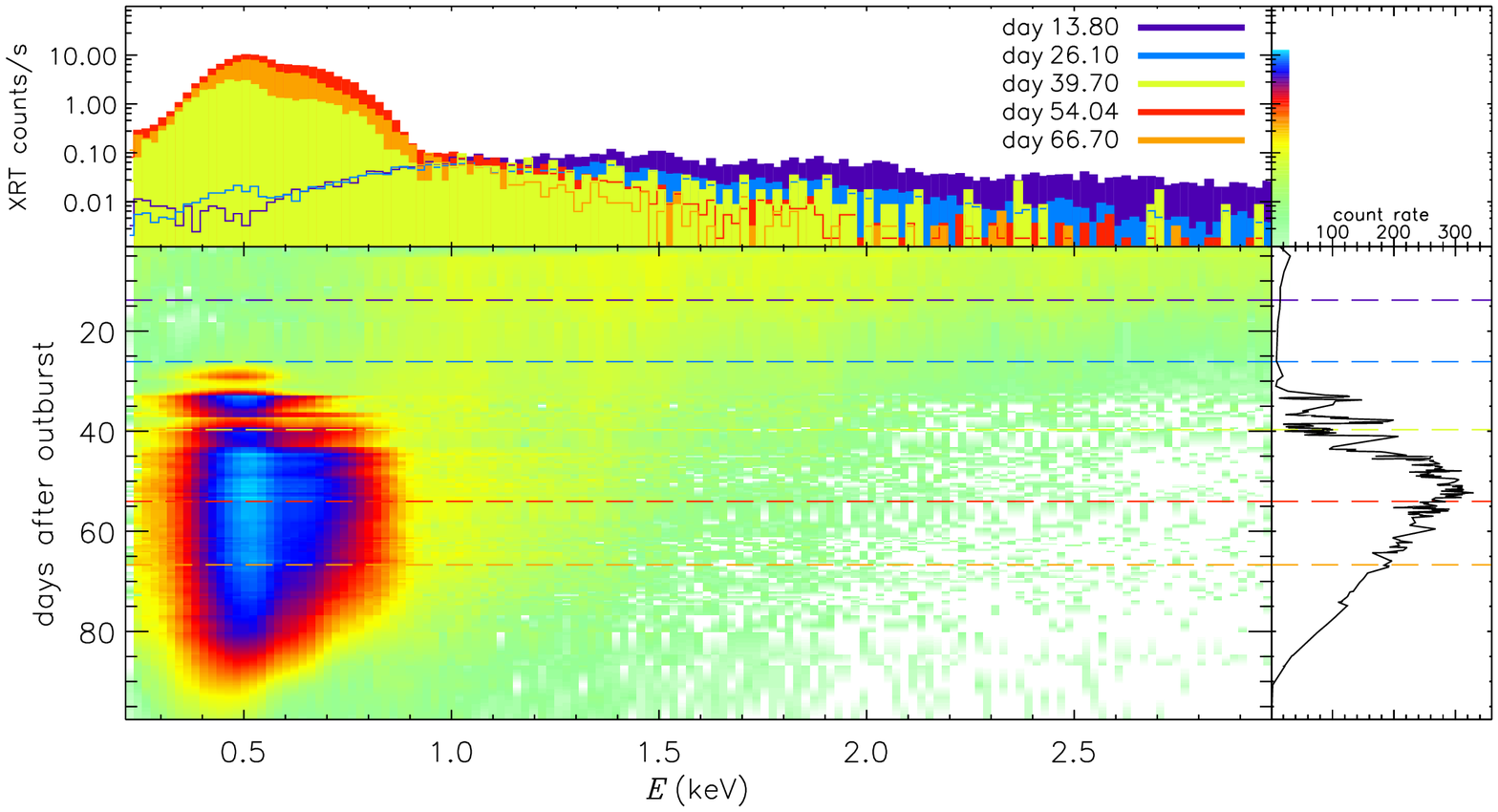}}

\resizebox{8.5cm}{!}{\includegraphics{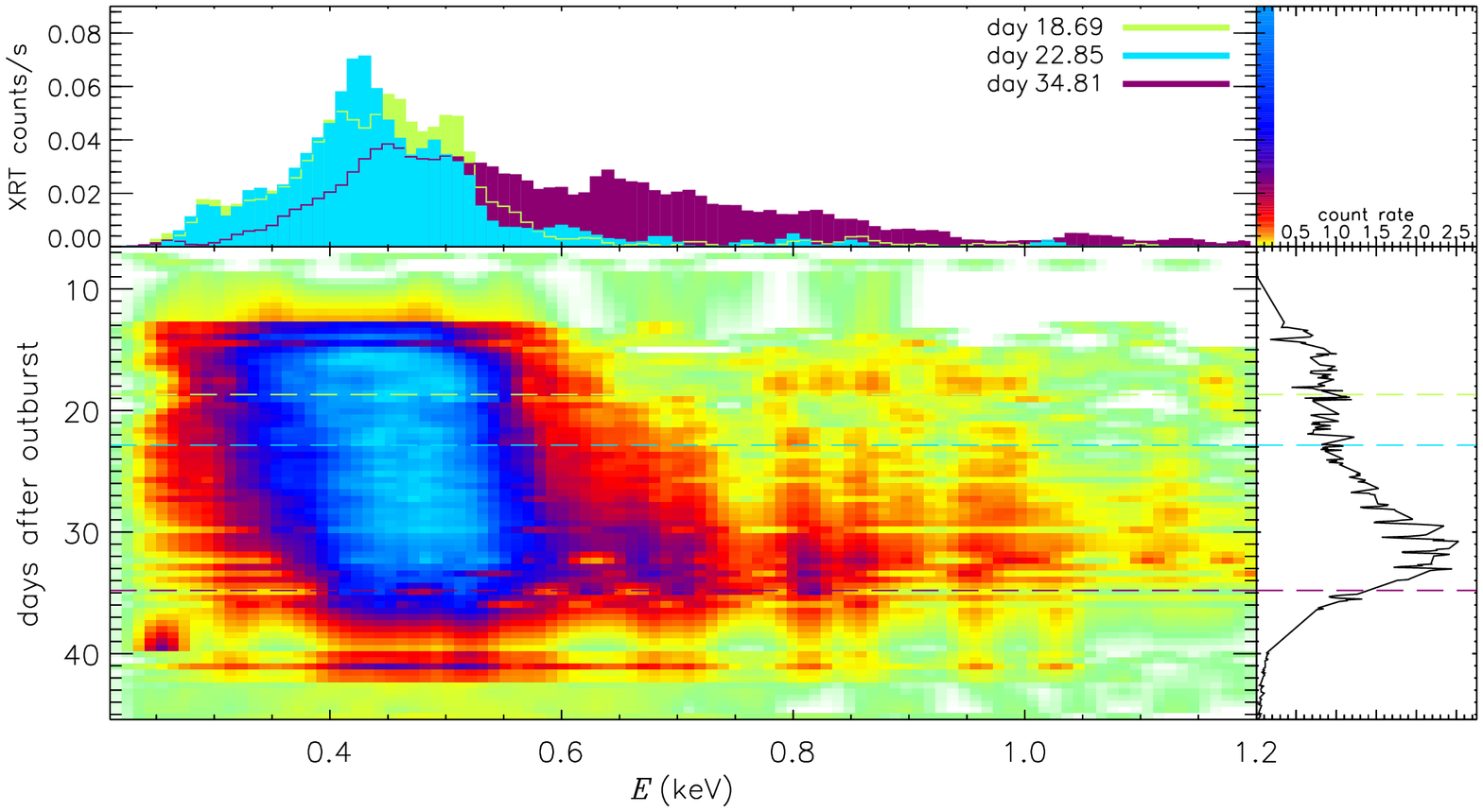}}

\end{centering}
\caption{\label{f:rsoph_swmap}Spectral evolution maps of the Swift
monitoring campaign of RS\,Oph (top) and U\,Sco (bottom). In each
plot, the top panel displays spectra extracted from time intervals
corresponding to the horizontal dashed lines of the same colours
in the time map below and in the light curve to the right
(rotated by 90$^{\rm o}$). The times when grating observations
were taken have been chosen. The
central panel illustrates the spectral time evolution with
elapsed observing time along the vertical axis and wavelength
across, on the same scales as the light curve in the right
and the spectra on top, respectively. The brightness is colour
coded, and for each colour, the corresponding count rate can be
derived from the colour bar in the top right.
{\small \it The data of RS\,Oph are the same as used by
\cite{osborne11} and have kindly been provided by Drs. Kim Page
and Andrew Beardmore of the University of Leicester}.
}
\end{figure}

The evolution of RS\,Oph during the rise phase into the
SSS phase led to some initial confusion as the nova seemed
to turn off quicker than expected but then increased again
in flux \citep{atel764}. Continued monitoring revealed
a high-amplitude variability phase during the early SSS phase
\citep{atel770}. RS\,Oph turned out not to be unique in having an early
variability phase as other nova monitoring observations displayed the
same. For example, in the slower nova V458\,Vul \citep{v458}, a
steep drop in count rate after the initial rise of SSS emission occurred
around day 410 (bottom right of Fig.~\ref{lc}). The on- and off states
during the early SSS phase were in fact strictly anticorrelated to the UV
\citep{schwarz2011}. The same phenomenon was seen in several other novae 
(see Fig.~\ref{lc} and \citealt{schwarz2011}), allowing the conclusion
that {\bf high-amplitude variations are a common phenomenon during
the early SSS phase of novae}.\\

Light curve modelling of the Swift light curve of RS\,Oph was
performed by \cite{hachisu07}, who assume that the optically
thick wind has stopped with the start of the SSS phase. The
high-amplitude variations are neglected in the light curve model.

During the SSS phase of RS\,Oph, periodic variations of $\sim 35$
seconds were present in the Swift and XMM-Newton X-ray light
curves \citep{atel770,atel801,ness_rsoph}. The same 35-s period was
also found in Swift observations of KT\,Eri, leading \cite{kteri35}
to conclude that it is not related to rotation but rather intrinsic,
e.g., nuclear burning instabilities. The complete set of
Swift X-ray spectra taken during the SSS phase of RS\,Oph were
studied in detail by \cite{osborne11}. The low-resolution Swift
spectra were modeled with the same
atmosphere code in which a set of fixed parameters was chosen
in addition to which the evolution of effective temperature, line-of-sight
absorption ($N_{\rm H}$), radius, and luminosity was studied.
As in the case of the hydro model used by \cite{vaytet2011}, a
highly complex model fitted to low-resolution spectra needs to be
constrained by high-quality grating spectra, if available,
however, currently no atmosphere model has so far been
found that reproduces the grating spectra in all details. For
example, \cite{nelson07} present a preliminary model using the
same code used by \cite{osborne11}, and it does not reproduce
the grating data. The assumed fundamental physics may thus be
wrong, but the approach is to consider only relative changes, and
\cite{osborne11} found that the early variability
phase was not only variable in brightness but also in
temperature and absorption.\\

A similar approach to find the underlying physical changes
that lead to spectral changes during the SSS phase were
presented by \cite{page09} using the monitoring data of
the fast nova V2491\,Cyg (see bottom left panel of Fig.~\ref{lc}).
Instead of an atmosphere model,
a simple blackbody fit was used, yielding the same parameters,
effective temperature, $N_{\rm H}$, radius, and luminosity,
again only inspecting the relative changes, even though the
underlying physics of the model are not fully correct.\\

The recurrent nova U\,Sco had been overdue to
explode again in 2009 \citep{schaefer_uscopred}
and was monitored in optical very closely until the
outburst was finally discovered in January 2010
\citep{usco_discovery}. It's particular importance
arises from the fact that the binary system is seen
edge on, yielding total eclipses \citep[e.g.,][]{schaefer_eclipse}.
Swift monitoring started immediately after
outburst, with the first observation only 1.2 days
after outburst. No detection in X-rays was found until
the onset of the SSS phase \citep{ATel2419}.
Consecutive Swift observations were centred around
predicted eclipse times, but with a series of short
snapshots, eclipses in X-rays were at most
suggestive from phase-folded count rates
\citep{ATel2442} from several cycles. A continuous
XMM-Newton observation indicated X-ray dips rather
than an eclipse (see Fig.~\ref{f:usco} in Sect.~\ref{s:swassist}). After
$\sim$day 30, the Swift count rates showed much clearer
eclipses, and a second XMM-Newton observation was
scheduled which then also contained a clear X-ray
eclipse. The entire Swift light curve is shown
in the third panel of the left column in Fig.~\ref{lc}
and the spectral evolution in the bottom panel of
Fig.~\ref{f:rsoph_swmap}. Clearly recognized can be
a conintuous increase in temperature with time.

\subsection{Pre-Swift XMM-Newton/Chandra observations}
\label{s:preswift}

The very first X-ray grating spectrum of a nova was
taken 2000, February 14 of V382\,Vel which was found
in the SSS state in a Chandra ACIS observation of 1999,
December 30 \citep*{burw02}. The resulting spectrum was
much fainter than expected, indicating that the nova had
turned off within only 2.5 months. The grating
spectrum was still useful as it displayed post-outburst
nebular emission lines that had never been seen before.
\cite*{ness_vel} inspected and analysed the data
using an emission measure distribution model. The line
profiles revealed double peaks indicative of different
kinematic components of the nebular plasma. Weak
continuum was seen that could arise from a
$3\times 10^5$-K WD, indicating that
it had not yet cooled down after the
outburst. Under the assumption of collisional
equilibrium in a low-density plasma as in stellar
coronae, the distribution of kinetic (electron)
temperature could be reconstructed from line
ratios. From the inferred temperature distribution and
systematic deviations of predicted and measured
lines fluxes, the elemental abundances could be
derived. Although the nova had gone through an
earlier so-called Fe-curtain phase in the UV
\citep{shore03}, the X-ray spectrum contains
no Fe lines. The abundance diagnostics confirmed
that Fe was at least 4\% underabundant with
respect to oxygen. \cite{ness_vel} argue that
instead of underabundant Fe, the elements that
produce the observed emission lines are overabundant.
Two more
Chandra observations of V382\,Vel were taken with
the ACIS, without a
grating, and fading emission lines were visible,
also with the lower CCD resolution \citep{burw02}.\\

\begin{figure}[!ht]
\resizebox{\hsize}{!}{\includegraphics{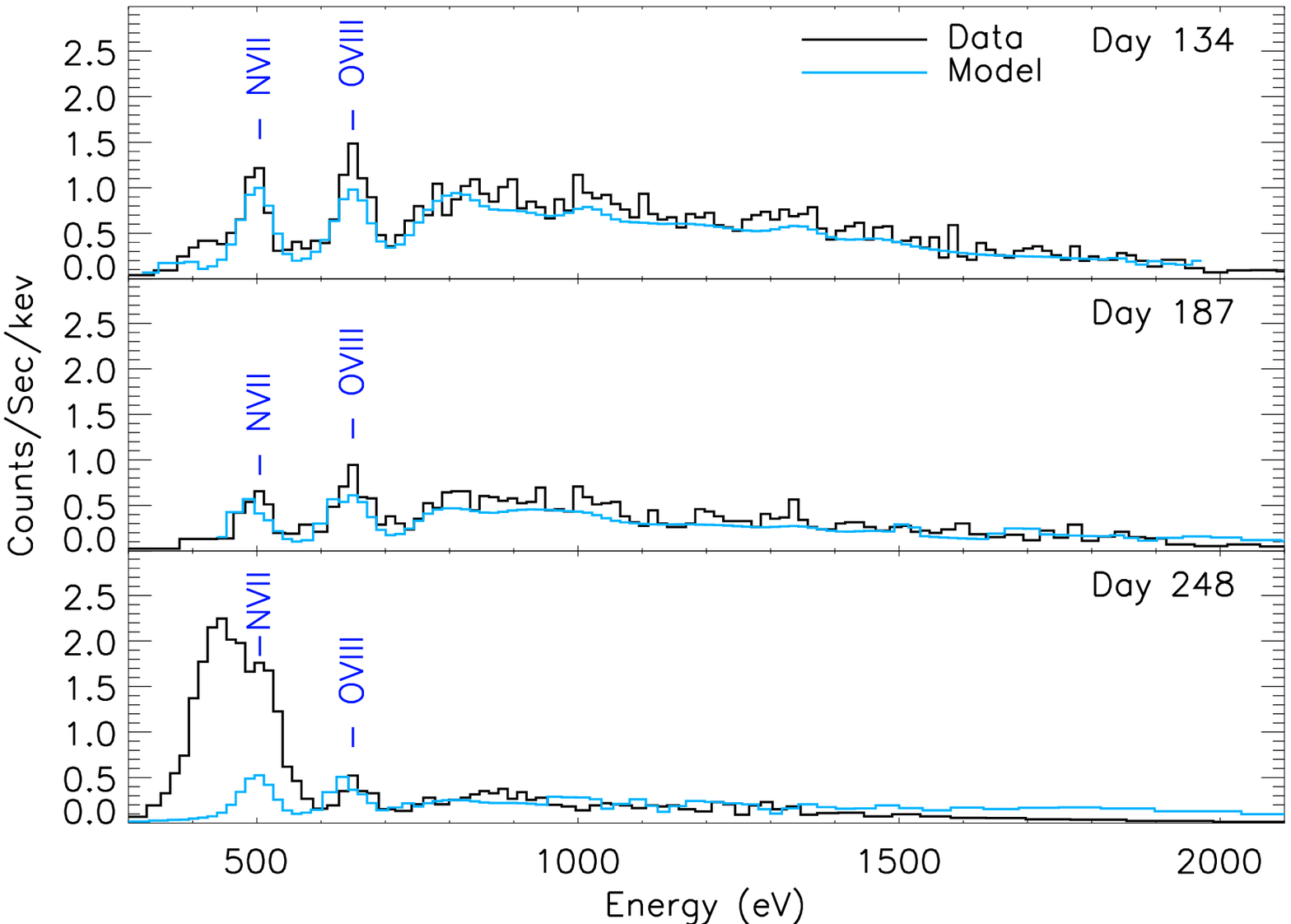}\includegraphics{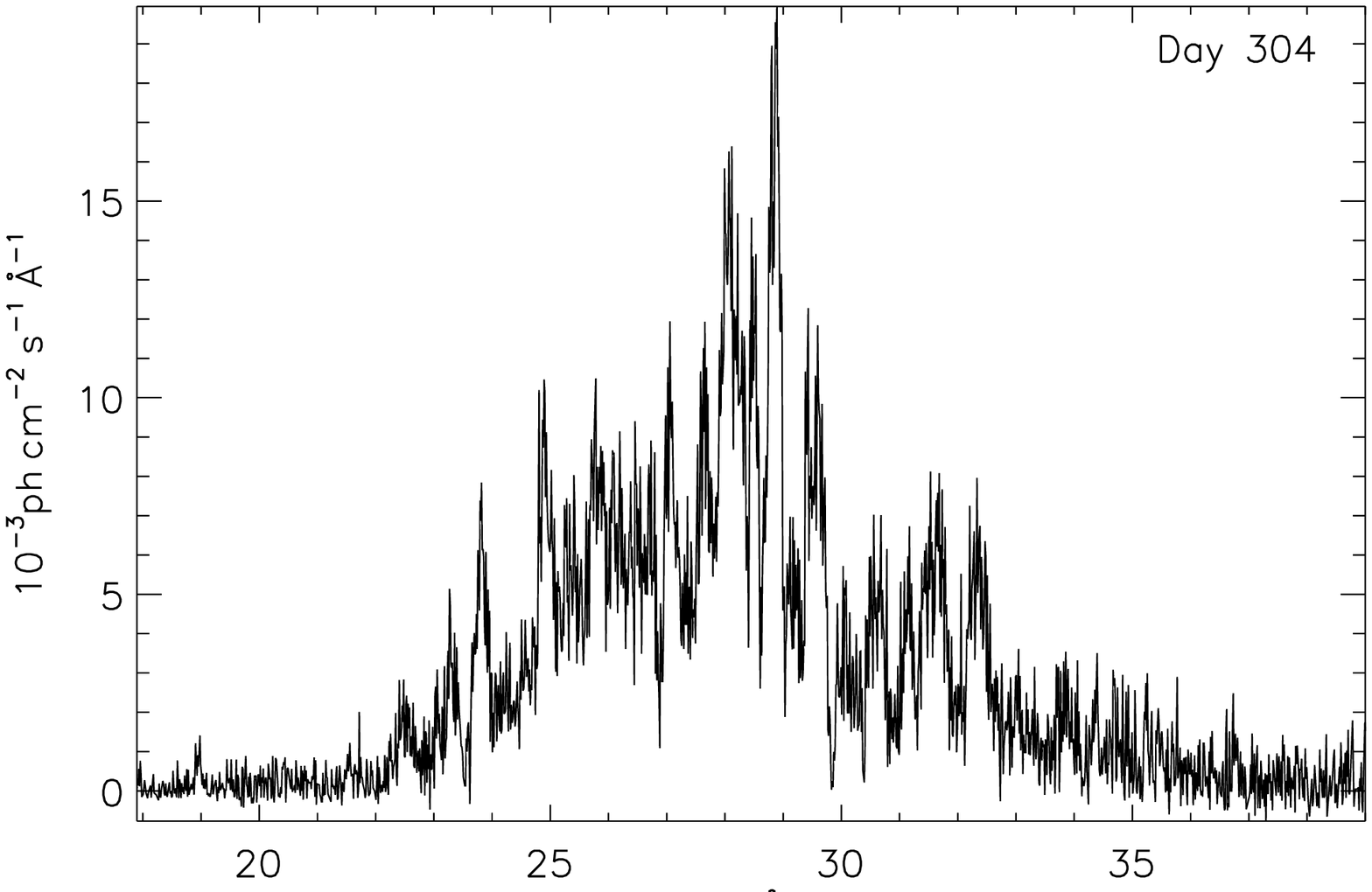}}
\caption{\label{f:v1494}Illustration of an observing strategy
without Swift. During the early evolution of V1494\,Aql, while it was
not clear what would be seen, short (5-6\,ks observations) with
the ACIS-I were taken (left panel; figure taken from
\citealt{rohrbach09}; for more details see there). The nova was still
too faint for the gratings but the faint pre-SSS ACIS spectrum
originated from optically thin thermal plasma that can be
parameterised by the electron temperature; see model plotted in
light blue. On day 248 after outburst, a clear SSS component
was detected, justifying longer grating observations that were
taken 2 months later (right panel).
}
\end{figure}

A few months later, the nova V1494\,Aql offered another opportunity
to study nova evolution with Chandra. Without any knowledge
how bright the nova may be in X-rays, a blind shot was taken
with the ACIS on day 134 after outburst, and the 5-ks
observation yielded a hard spectrum with two resolved emission
lines but no signs of SSS emission (see Fig.~\ref{f:v1494}).
Fortunately, significant X-ray emission was present already
before the SSS phase started, so more observations in $\sim 3$
months intervals yielded secure results with good perspectives
for eventually finding SSS emission. On day 248, clear indications
for SSS emission arose, albeit still not quite as bright as
in other novae. On day 300, the grating was then employed,
yielding the first SSS spectrum of a nova in high resolution.
This spectrum is different than expected, showing continuum
plus emission lines. 4 days later, a deeper exposure was
taken, yielding a similar spectrum, thus the SSS spectrum
is not an atmospheric spectrum as expected. It may be related
to the high system inclination angle \citep{v1494_eclipse}
(Ness et al. in preparation, and Fig.~\ref{f:speccat}).
To this date,
no satisfactory spectral model could be found. For discussion
also on timing analysis, I refer to \cite*{rohrbach09}. Without
any information how long the SSS phase may last, the next
observation was taken one year later, but the nova had already
turned off, yielding a weak detection only in the zero-th
order of the grating observation. 5 years later, Swift observed
again with no further detections (Sect.~\ref{s:swmonitoring}).\\

\begin{figure}[!ht]
\begin{centering}
\resizebox{8.5cm}{!}{\includegraphics{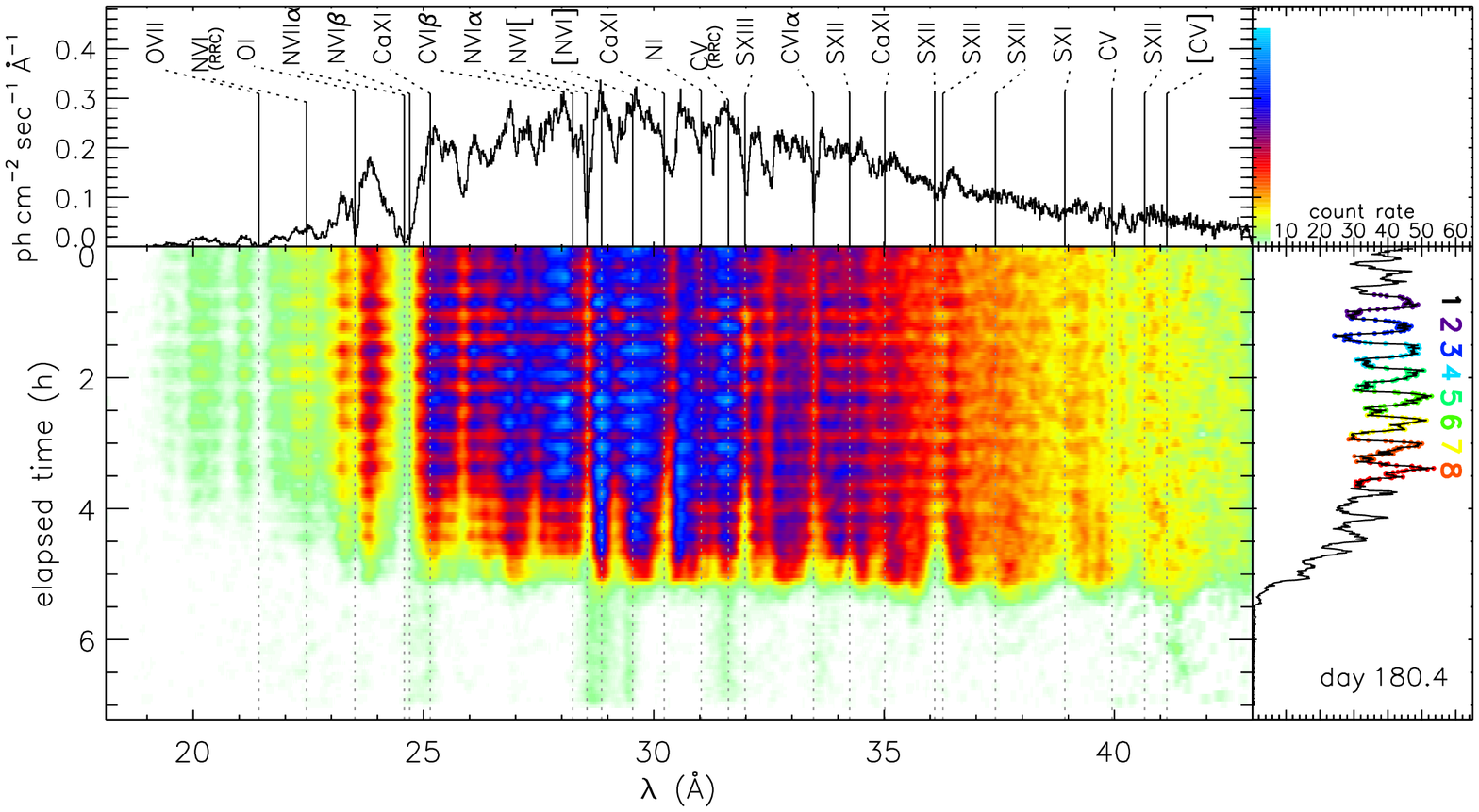}}

\resizebox{8.5cm}{!}{\includegraphics{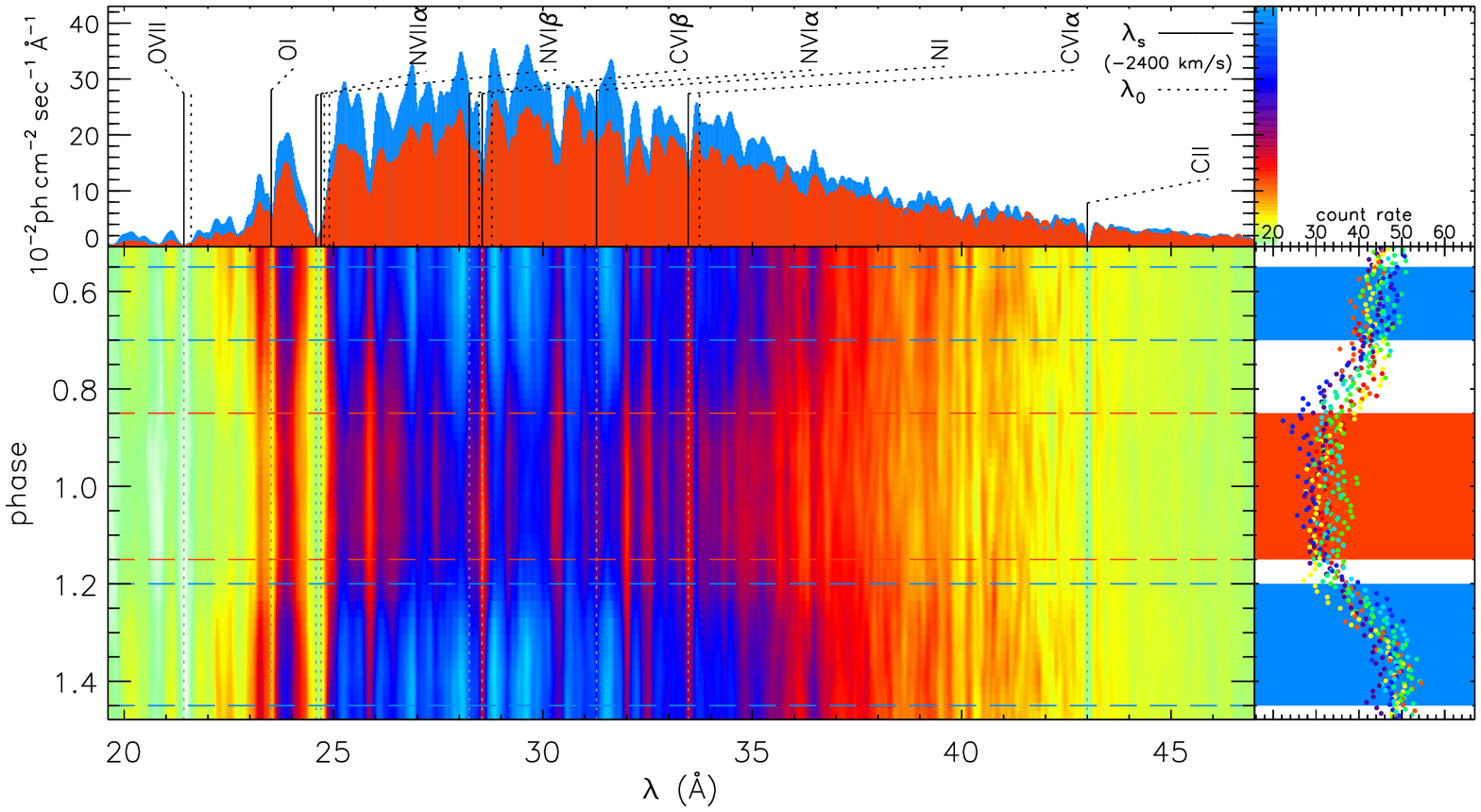}}

\resizebox{8.5cm}{!}{\includegraphics{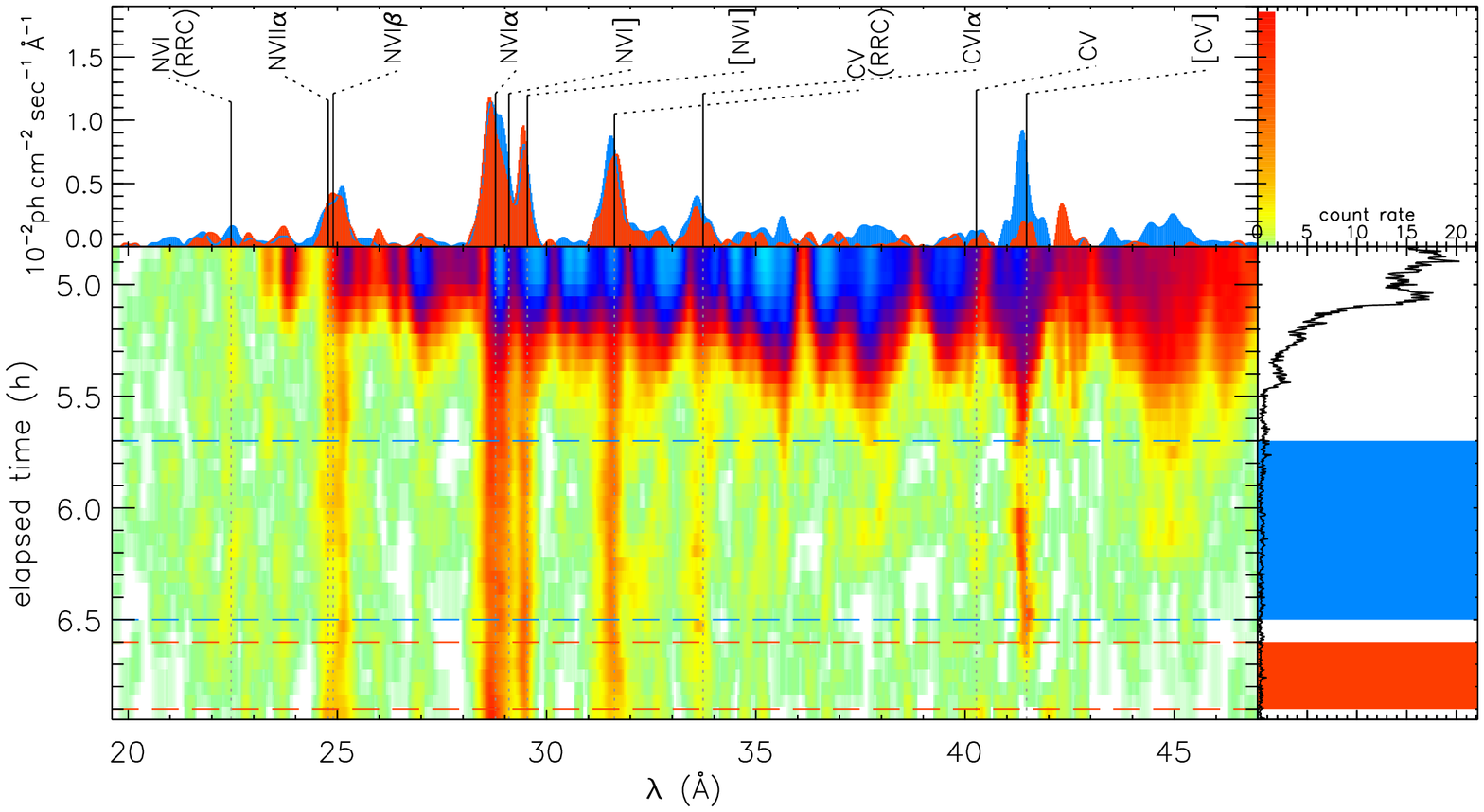}}

\end{centering}
\caption{\label{f:v4743maps}Spectral evolution maps of a Chandra
observation of V4743\,Sgr on day 180.4 after outburst, each plot
illuminating a different aspect using the same illustration concept
as explained in Fig.~\ref{f:rsoph_swmap}.
In the top, the evolution over the entire observation is shown.
After a phase of periodic variations (period 1313.6\,seconds, eight clean
cycles are marked), a steep decline sets in, leaving behind weak
emission lines. The spectrum can be seen to first fade at shorter
wavelengths. Spectral changes along the periodic variations are
illustrated by the phase-folded plot in the middle, folding the eight
cycles marked in the top. The colours of the plot symbols in the right
indicate from which cycle the corresponding data points originate.
In the bottom, the transition from bright to faint is illustrated.
The forbidden line of C\,{\sc v} at $\sim 41.5$\,\AA\ and some broader
features around 37\,\AA\ and 45\,\AA\ seem to fade with time.
}
\end{figure}

Also the nova V4743\,Sgr (2002) had to be studied without
additional information from other X-ray observatories. Again,
in order to avoid non-detections in a long grating observation,
a first, short, 5-ks, snapshot was taken in November 2002,
two months after outburst. Again, a clear detection was
found but no SSS spectrum. The nova was then not visible
for 4 months, but as soon as it came back from behind the
sun, a grating spectrum was taken (180.4 days after outburst),
yielding the first really bright SSS spectrum with continuum
and deep absorption lines from highly ionised species \citep{v4743}.
A graphical illustration of the spectral evolution during this
observation is shown in Fig.~\ref{f:v4743maps}.

The observation starts with high-amplitude periodic variations,
and eight clean cycles are marked with different colours
in the light curve, turned around by 90$^{\rm o}$, in the
right panel of Fig.~\ref{f:v4743maps}. These eight cycles have
been used to create a phase-folded light curve and corresponding
spectral map, shown in the middle plot. The two spectra in the top are
integrated over all phase minima and phase maxima, respectively,
showing that the minima spectra are systematically softer.
These periodic X-ray variations have been seen in all
X-ray observations, including those taken post outburst.
Detailed timing analyses have been carried out by
\cite{leibowitz06,dobrness09}, discovering the double nature
of the frequency during the SSS phase which later merges
into a single frequency \citep{dobrness09}.

About 2 hours before the end of the observation, the count
rate suddenly dropped from 40-60 counts per second to nearly
zero within very short time. No technical fault was identified,
but a physical cause for such a steep decline was not easily
found. At that time, the high-amplitude variations during the
early SSS phase described above in Sect.~\ref{s:swmonitoring}
had not yet been discovered, and it appears plausible that this
decline was in fact the first time this phenomenon was seen.

In the top plot, one can see that the harder part of the
spectrum, shortward of $\sim 28$\,\AA, started earlier with the
decline than the softer part, and that emission lines were left
behind, both reported by \cite{v4743}. The evolution of the
faint emission lines after the decline is illustrated in the
bottom, and one can see that some emission lines are variable.
The forbidden line of C\,{\sc v} (1s$^2\,^1$S$_0$--1s2p\,$^3$S$_1$)
at 41.47\,\AA\ is much stronger than the
1s$^2\,^1$S$_0$--1s2p\,$^1$P$_1$ resonance line at 40.27\,\AA.
While at about 5.7 hours after the start of the observation, the
$[${C\,{\sc v}$]$ line disappears, it reappears between 6.0 and
6.6 hours, at first red shifted and then gradually changing
to being blue shifted. If the $[${C\,{\sc v}$]$ emitting material
originated from plasma that co-rotates with the WD 
at 2400\,sec spin period, red/blue shifts of $\pm 0.15$\,\AA\
implied an orbital radius of 3\,R$_\odot$. It bears closer
investigation, however, why a plasma so far away would not
produce any resonance line emission.

The sudden decline triggered an XMM-Newton observation only two
weeks later (day 196.1 after outburst), in which the source was
as bright as before the decline started \citep{v4743_xmm},
indicating that it was a transient phenomenon, strengthening
the link to the early high-amplitude variability phase (see
Sect.~\ref{s:swmonitoring}). Denser monitoring than
for V1494\,Aql was chosen with three more observations on days
301.9, 371, and 526.1 after outburst, and all yield the same
general spectrum with some changes in temperature and
in several details, but no further decline as on day 180.6 was
seen.\\

All five SSS spectra of V4743\,Sgr were modeled by \cite{rauch10}
using the T\"ubingen Model Atmosphere Package (TMAP), a non-LTE
atmosphere model that is optimised for hot
WDs. The evolution of the effective temperature was
described to have been reached on day 196.1, after which it
remained stable at least until day 371. A massive WD
of 1.1-1.2\,M$_\odot$ was deduced, and the photosphere was
found to be nitrogen-rich and carbon-deficient, indicative
of CNO cycle processed material. Under the assumption that
the observed photospheric emission arises from the surface of
the WD, it was concluded that some of the accreted material
was not ejected and remained instead on the WD.
Towards later observations, a reduced nitrogen abundance was
derived, implying that new, N-poorer material, was present
that may come from the companion via renewed accretion.\\

These conclusions may depend to some degree on the assumptions
that, in spite of the complexity of the TMAP code, must still
be considered somewhat simplified. \cite{ness_v2491} argued that
assumptions such as hydrostatic equilibrium and a plane parallel
geometry are only valid in isolated hot WDs, for which the TMAP
code is designed, but not in expanding nova ejecta. The blue
shifts of the atmospheric absorption lines in the X-ray grating
spectra give direct whitness of the continued expansion and thus
a non-static environment. In the
TMAP model, \cite{rauch10} introduced an artificial blue shift
to better reproduce the data, but the entire atmosphere
structure may be fundamentally different in an expanding
environment. Attempts to model also the expansion in non-LTE
radiative transport calculations have been made in the PHOENIX
code (see, e.g., \citealt{hauschildt92,phx_expand}).
Clearly, the complexity increases even further, but the X-ray
grating spectra are of sufficient quality to use such approaches.
The PHOENIX code was originally designed to model UV spectra,
but has later been expanded to the X-ray regime \citep{petz05}
and improved by \cite{daanthesis} with first results shown by
\cite{vanrossumness09} who found that atmosphere models that
account for the expansion yield vastly different results from
static models. Most recently, \cite{vanRossum2012} presented a
full description of a new model, based on PHOENIX, called
"Wind-Type" (WT) SSS Nova Model, that will be made publicly
available. The same five grating spectra of V4743\,Sgr have been
modeled, yielding different principal parameters than those
found by \cite{rauch10}. The effective temperatures are generally
lower while $N_{\rm H}$ is higher. The values found by
\cite{vanRossum2012} are in better agreement with Galactic neutral
hydrogen maps of \cite{Kal05} and \citet{dickey90} as
implemented in the HEASARC Galactic $N_{\rm H}$
tool\footnote{http://heasarc.nasa.gov/cgi-bin/Tools/w3nh/w3nh.pl}
yielding $1.1\times10^{21}$ and $1.4\times10^{21}$\,cm$^{-2}$,
respectively. The differences between \cite{vanRossum2012} and
\cite{rauch10} can partly be explained by a different approach
in determining $N_{\rm H}$, but the main difference is caused
by two additional parameters, $v_\infty$ and $\dot{M}$ in
\cite{vanRossum2012}, parameterising the expansion. More work
will be needed to investigate the effects of non-solar
abundances. Well constrained model parameters will guide
theoretical nova evolution models. The low-resolution spectra
taken during Swift monitoring can be used to interpolate
the models that have been constrained in detail by 2-3 grating
spectra.

\subsection{Swift-assisted XMM-Newton/Chandra observations}
\label{s:swassist}

The greatest difficulty in getting grating spectra during
the SSS phase is finding the right moment. This has been
overcome with the use of Swift to get short snapshots
over much shorter time intervals (see Sect.~\ref{s:swmonitoring}).
The Swift count rates were used to calculate the
required grating exposure times.\\

The first time that the potentials of Swift for nova
studies were systematically explored was during the 2006
outburst of the recurrent, symbiotic, nova RS\,Oph.
Initial Swift observations established an extremely high
X-ray brightness level, with severe pile up in the
photon counting mode. There was no doubt that short
grating observations would be highly informative, and
simultaneous XMM-Newton and Chandra observations were
organised to take place only two weeks after outburst.
While Swift monitoring continued, grating observations
were taken whenever significant changes in brightness
and spectral shape were seen with Swift, yielding a
total of three XMM-Newton and three Chandra observations
before the end of the SSS phase (see 2nd panel in
left row of Fig.~\ref{lc}). The
nova remained interesting also after it turned off,
and more grating observations were taken. An overview
of the grating observations was given by \citet{rsophshock}
which is reproduced in Fig.~\ref{f:rsoph}. The first two
spectra were extracted from the two grating arms of the
Chandra/HETGS, the Medium Energy Grating (MEG) and the
High Energy Grating (HEG). These observations were
taken simultaneously with the XMM/RGS (third
spectrum in Fig.~\ref{f:rsoph}).\\

The top four spectra in Fig.~\ref{f:rsoph} were taken
during the shock phase,
before the SSS phase started while, in the symbiotic
system, the nova ejecta ran through the dense
stellar wind of the giant companion star, dissipating
significant amounts of kinetic energy during shocks
that was partially converted to X-ray emission.
The X-ray spectrum is that of a plasma in collisional
equilibrium, evidenced by line ratios of He-like
emission lines of Mg \citep{nelson07}. Two different groups
have analysed the same spectra in different ways:
\cite{nelson07} applied an {\tt xspec} \citep{xspec}
multi-temperature model with solar abundances,
needing 4 temperature components.
\cite{rsophshock} used two approaches, first
the same as \cite{nelson07} but with variable
abundances, needing then only three temperature
components. High overabundance of Mg and Si
and underabundance of Fe, relative to oxygen, were found.
Since a plasma
of three distinct temperature components is not
necessarily physically realistic, \cite{rsophshock}
also applied the same method already used for
V382\,Vel assuming a continuous temperature
distribution (see above and \citealt{ness_vel}) and
found the same general abundance trends but with
better constraints of N and S. Comparison of the
abundances with those in single giants revealed
considerable anomalies suggesting that the companion
in RS\,Oph has changed substantially by the
continued mass loss via accretion. Alternatively,
these abundances would reflect the composition of the
WD which would question the system
to be a SN\,Ia progenitor which requires a
carbon-oxygen rich (CO) WD while
the observed abundances in the X-ray plasma
would indicate more an oxygen-neon (ONe) white
dwarf. The difference in nuclear binding energy
between $\alpha$ elements (such as neon, magnesium,
silicon) and iron is not sufficient to disrupt
the WD while nuclear fusion chain reactions
starting from carbon, as found in a CO WD, produce
more energy.\\

\begin{figure}[!ht]
\resizebox{\hsize}{!}{\includegraphics{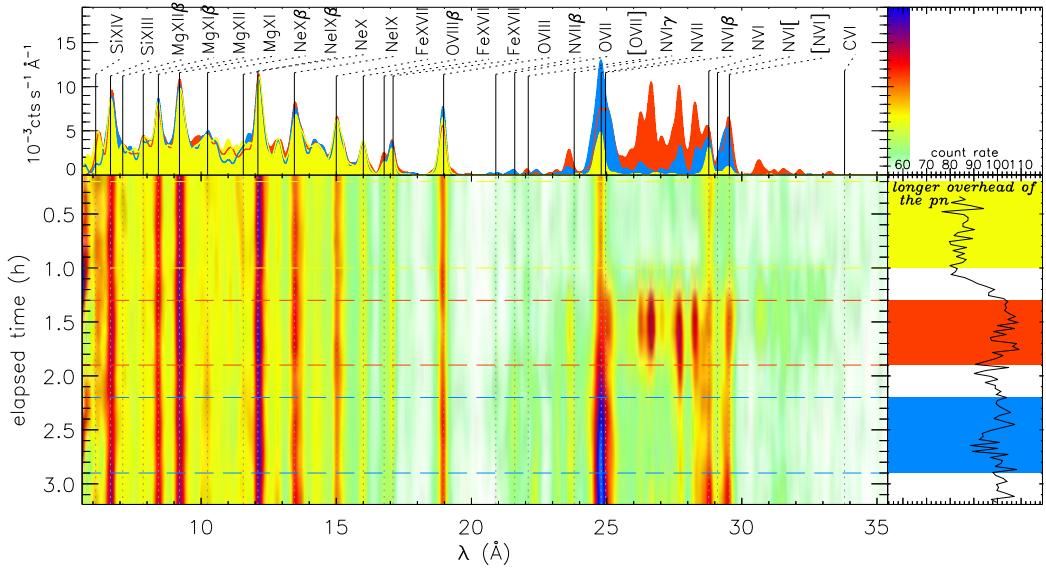}}
\caption{\label{f:rsophmap_early}Spectral evolution map
of XMM-Newton observations of RS\,Oph on day 26.1 (see
Fig.~\ref{f:rsoph_swmap} for description of the illustration
concept). While the spectrum
shortward of 20\,\AA\ remains constant, the rise in total
flux about one hour after the start of the observation
is owed to a new soft component between 24-31\,\AA, first
described by \cite{nelson07}.
The spectrum of the new soft component is not typical SSS
emission but consists of unknown emission lines.
}
\end{figure}

The SSS phase of RS\,Oph may have started during
the XMM-Newton observation taken on day 26.1 in
which \cite{nelson07} observed the appearance of a new
soft component. While the high-energy part of the spectrum
(fourth panel in Fig.~\ref{f:rsoph}) could be
modelled with the same approaches of collisional
plasma as for day 13.8 \citep{rsophshock}, the new
soft component between $\sim 24-31$\,\AA\ coincides
in wavelength with the soft tail of the later SSS
phase (see fourth and fifth panels in
Fig.~\ref{f:rsoph}). In a low-resolution spectrum,
the new component would have been interpreted as first
signs of atmospheric emission, and a blackbody fit
to the simultaneous XMM-Newton/MOS spectrum yields
indeed a good fit. However, in the high-resolution RGS
spectrum, several unknown overlapping emission lines
can be seen, quite atypical of SSS emission. An
illustration of the time evolution of the appearance
of the new soft component during the observation is
given with Fig.~\ref{f:rsophmap_early}. Only two
lines of N\,{\sc vi} (1s-2p triplet) at 28.8-29.5\,\AA\
and N\,{\sc vii} (1s-2p doublet) plus N\,{\sc vi} (1s-3p)
at $\sim 25$\,\AA\ can be identified while all other
features are of unknown origin. \cite{nelson07}
speculate these may be highly blue-shifted 1s-3p and
1s-4p lines of C\,{\sc vi}, but the principle
C\,{\sc vi} 1s-2p line at 33.8\,\AA\ is not present
(see Fig.~\ref{f:rsophmap_early}), and also the C
abundance was argued in other context of the same
article to be extremely low. From Fig.~\ref{f:rsophmap_early}
it can be seen that the unknown emission features
were present for only 40-50 minutes (between the
two horizontal dashed red lines), suggesting these
may be instable transitions, e.g., between excited states.
A conclusive interpretation of the day-26.1 spectrum
of RS\,Oph is still outstanding but important as it may hold
clues of the transition into the SSS phase.\\

A few days later, Swift observed a steep
rise into the SSS phase, and a Chandra LETG observation
was scheduled which fell, however, into an unanticipated
low state that occurred soon after. Nevertheless, an
extremely well exposed spectrum was obtained. Continued
Swift monitoring allowed scheduling of two more grating
observations during the peak and decline of the SSS
phase (see Fig.~\ref{lc}). The time evolution of this
observation is illustrated in Fig.~\ref{f:rsoph_grmap}.
Also on shorter time scales, the nova was highly
variable. A 1000-s delay between brightness and hardness
light curves in the day 39.7 (see right panel of
Fig.~\ref{f:rsoph_grmap}) observation led
\cite{ness_rsoph} to conclude that photoionisation of
neutral oxygen led to variations in transparency,
allowing more (harder) X-ray light to escape when oxygen
is ionised. Considerable reductions in the depth of the
neutral oxygen edge towards the later observations
support this conclusion.

\begin{figure}[!ht]
\resizebox{\hsize}{!}{\includegraphics{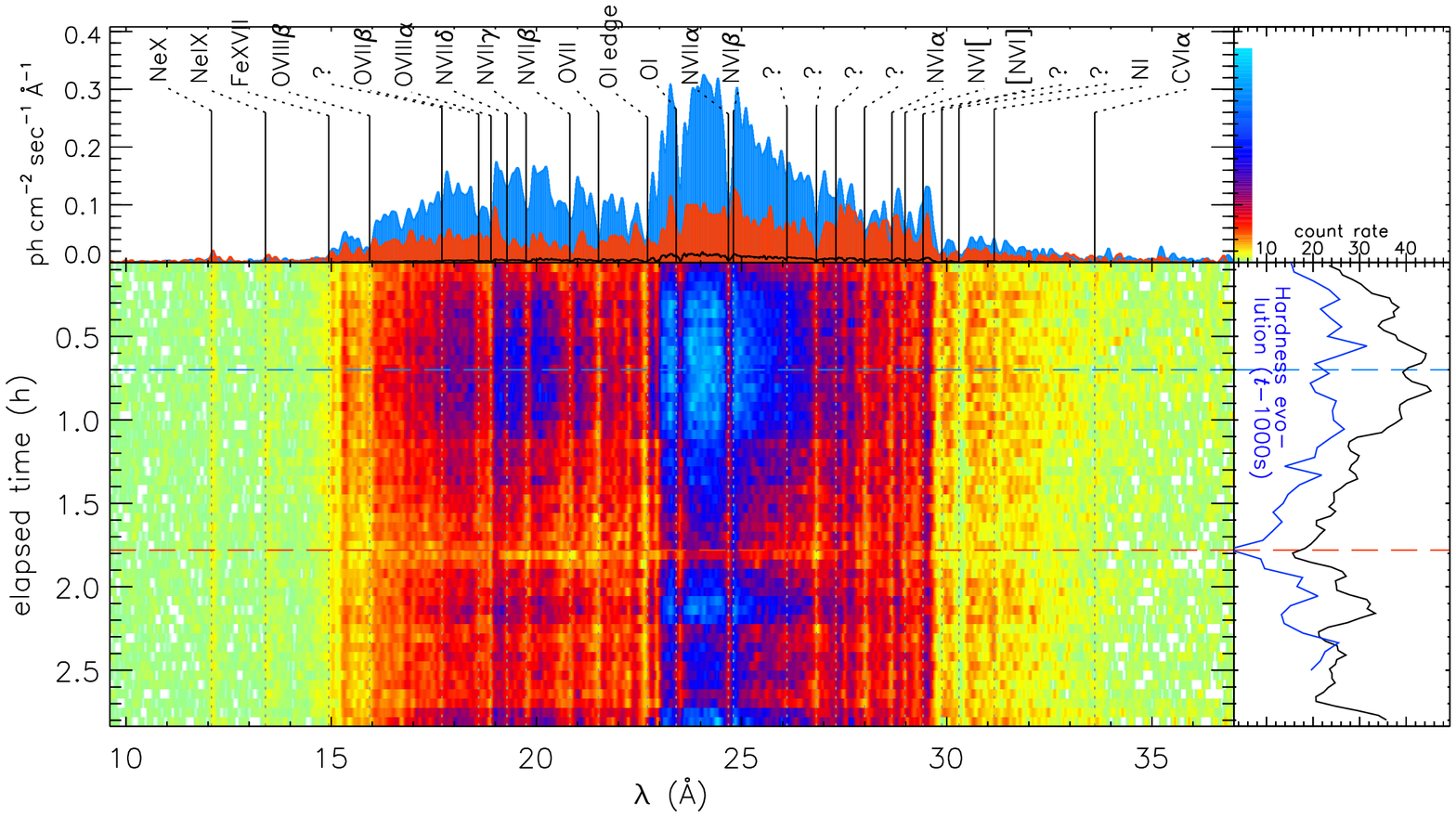}}
\caption{\label{f:rsoph_grmap}Spectral evolution map, as
explained in Fig.~\ref{f:rsoph_swmap}, for the Chandra
observation taken during the high-amplitude variability
phase of RS\,Oph, on day 39.7. A 1000-second shift of
the hardness light curve versus brightness discovered by
\cite{ness_rsoph} is illustrated in the right by the
shifted hardness light curve in blue.
}
\end{figure}

\begin{figure}[!ht]
\resizebox{13cm}{!}{\includegraphics{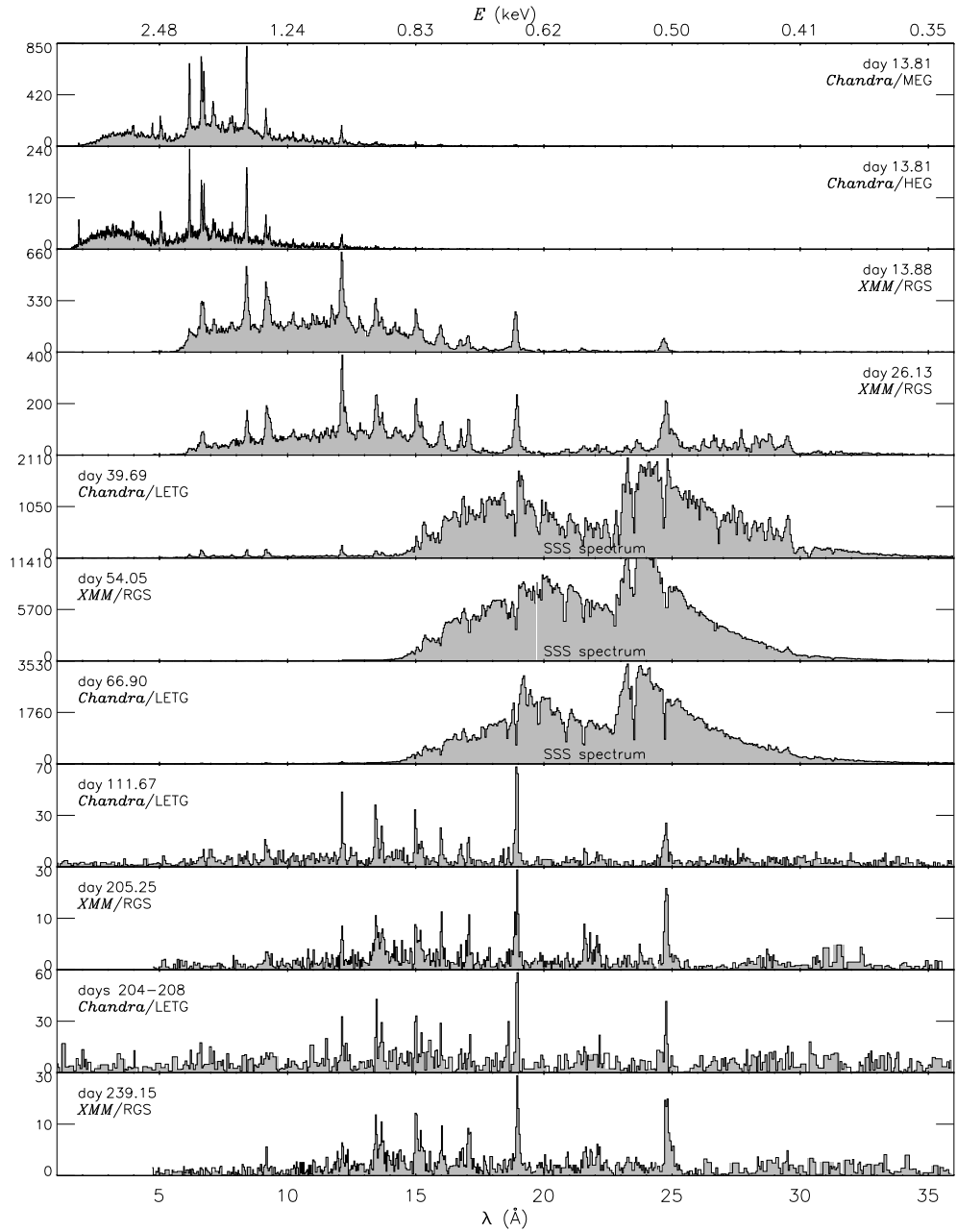}}
\caption{\label{f:rsoph}All grating observations taken
during the 2006 outburst of RS\,Oph (figure taken from
\citealt{rsophshock}).
}
\end{figure}

Based on the Swift light curve of the famous recurrent nova
U\,Sco, shown in the third panel of the left column in
Fig.~\ref{lc} and the bottom plot in Fig.~\ref{f:rsoph_swmap},
an initial Chandra observation was scheduled on day 18.7 after
outburst but was not centred around eclipse. Later,
on day 22.9, a longer continuous XMM-Newton observation
was scheduled that resolved the details of the
X-ray brightness during optical eclipses.
Guided by the contemporaneous OM light curve that
clearly showed an eclipse, \cite{ness_usco} showed that dips
occurred between quadratures (phases -0.75 to 0.25; see
left panel in Fig.~\ref{f:usco}).
A few days later, the Swift count rates showed clear
eclipses, and a second XMM-Newton observation was
scheduled to resolve the details (see right panel in
Fig.~\ref{f:usco}). The X-ray spectrum
during the SSS phase was atypical again, with strong
emission lines on top of blackbody-like continuum,
resembling the proto typical SSS Cal\,87 in the LMC,
which is also an eclipsing system. \cite{ebisawa10}
discussed the possibility that the accretion disk in
Cal\,87 blocks the central emission from the WD,
and all observable X-ray emission originates from
Thomson scattering within the Accretion Disk Corona
(ADC). Additional resonance line scattering can produce
the observed emission lines. To transfer these
conclusions to U\,Sco, an accretion disk must already
have been present or in the process of re-establishment.
The dips in X-rays can be explained by neutral clumpy
material within an accretion stream from the companion
star. Neutral gas is transparent to UV/optical light
but opaque to X-rays. Once the accretion disk has flattened,
no more dips are seen in X-rays.\\

\begin{figure}[!ht]
\resizebox{\hsize}{!}{\includegraphics{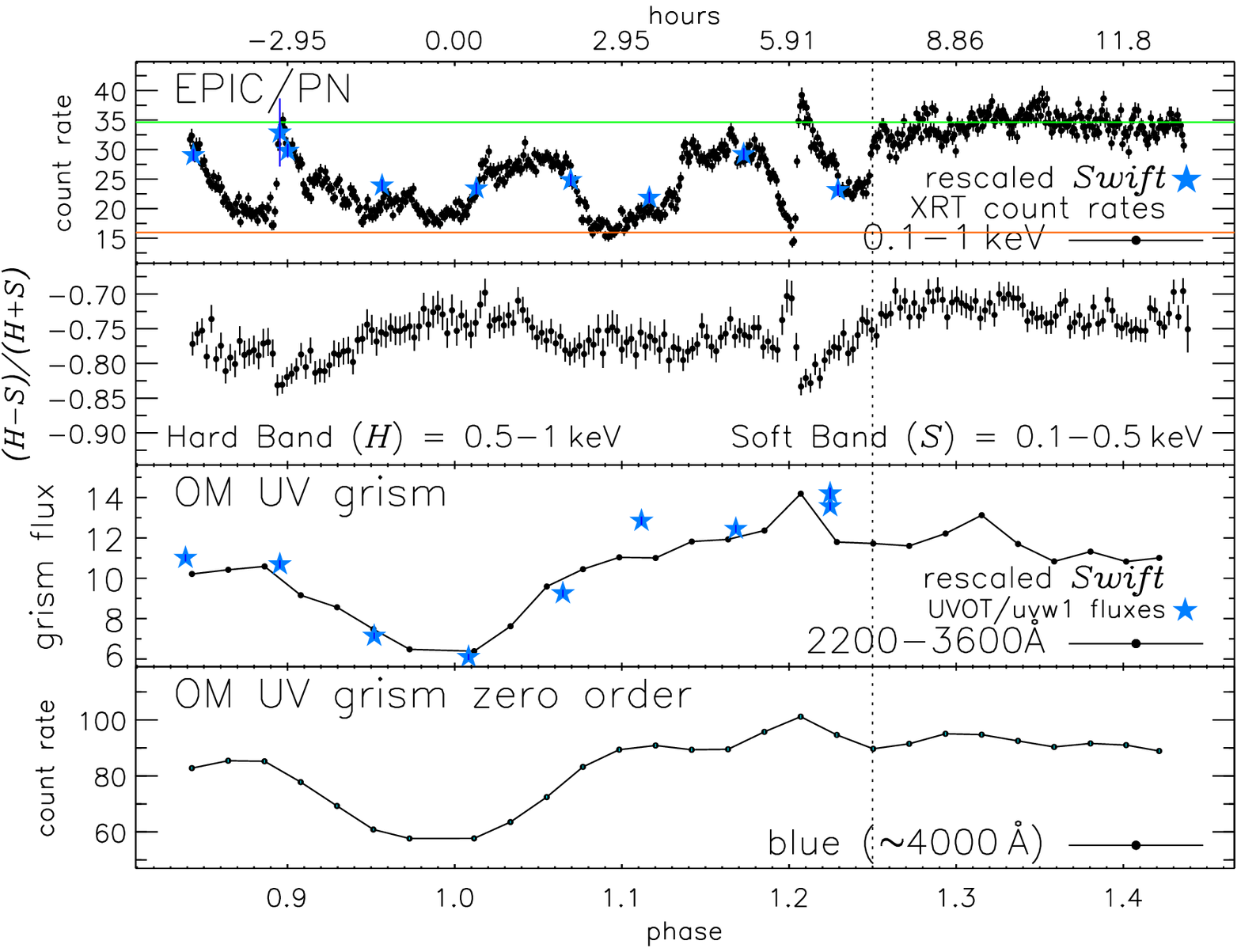}\includegraphics{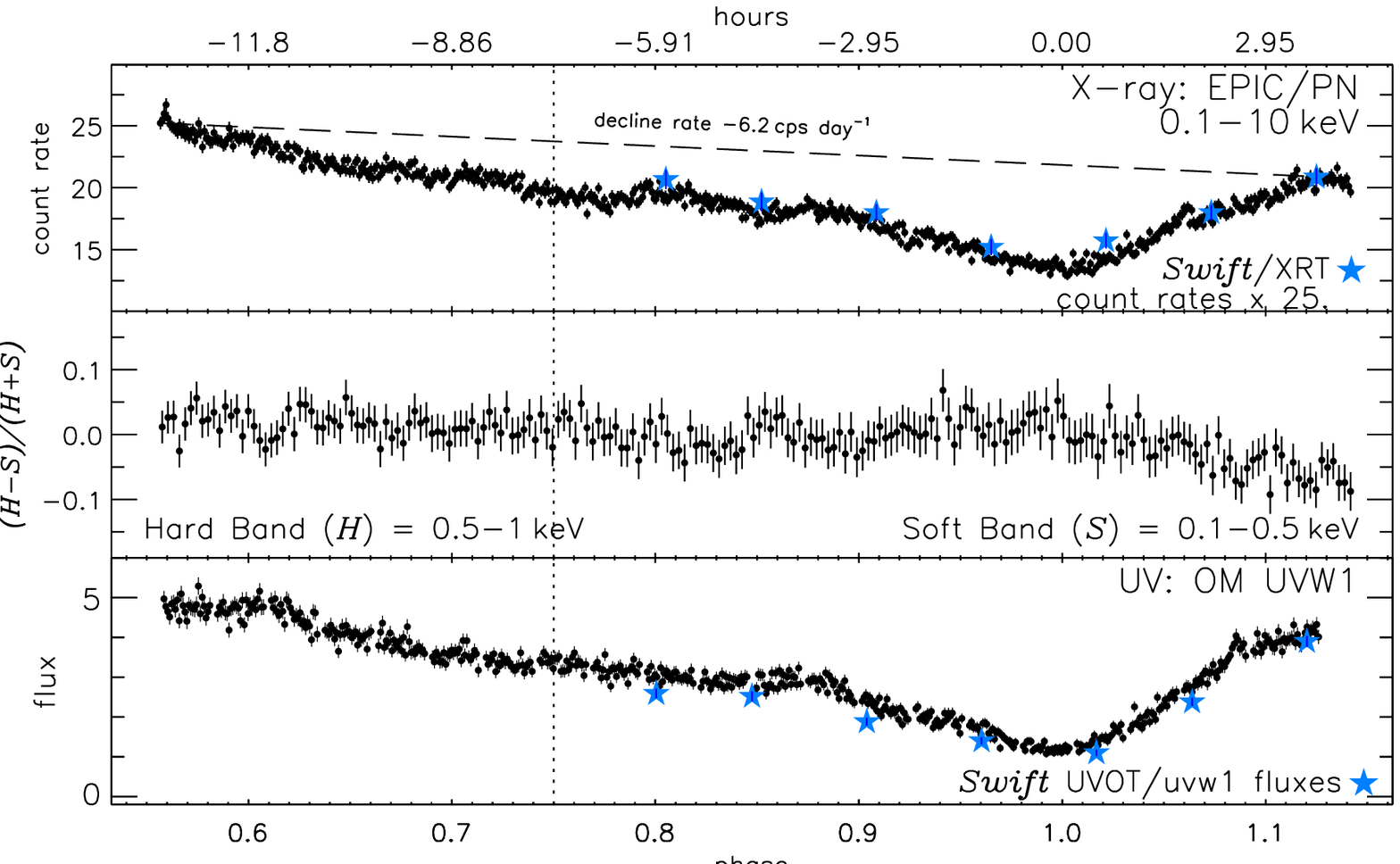}}
\caption{\label{f:usco}XMM-Newton X-ray and UV light curves of U\,Sco,
centred on eclipses (taken from \citealt{ness_usco}).
In the earlier observation (left), X-ray dips and a clean
eclipse in UV and optical are seen, while in the later
observation (right), also the X-ray light contains a clean
eclipse. \cite{ness_usco} interpret these changes as signature
of the re-establishment of accretion during the SSS phase.
}
\end{figure}

The very fast nova V2491\,Cyg entered the SSS phase just
a bit more than a month after outburst. Guided by the Swift
snapshots (see bottom left panel in Fig.~\ref{lc}), two
XMM-Newton observations were triggered in short succession,
on days 39.9 and 49.6, both extremely well exposed. Without
Swift, this quick evolution could not have been followed.
Significant X-ray variability was found on day 39.9
(\citealt{ness_v2491}, see Fig.~\ref{f:v2491cyg_smap}).
The absorption lines were highly blue shifted indicating
high expansion velocities. \cite{ness_v2491} also showed
that currently publicly available atmosphere models are
not sufficient to reproduce these high-quality spectra.
Preliminary PHOENIX atmosphere models to the V2491\,Cyg
spectra were presented by \cite{vanrossumness09}, using
a combination of a hydrostatic core and a hydrodynamic
expanding envelope. The additional expanding envelope
yields a clearly better fit to the data, but the large
majority of absorption lines are not reproduced.
\cite{pinto12} developed a new approach using a multicomponent
SPEX \citep{spex} model, consisting of emission and absorption
components, but also this approach did not reproduce the
spectra in a formally satisfactory way.\\

\begin{figure}[!ht]
\resizebox{\hsize}{!}{\includegraphics{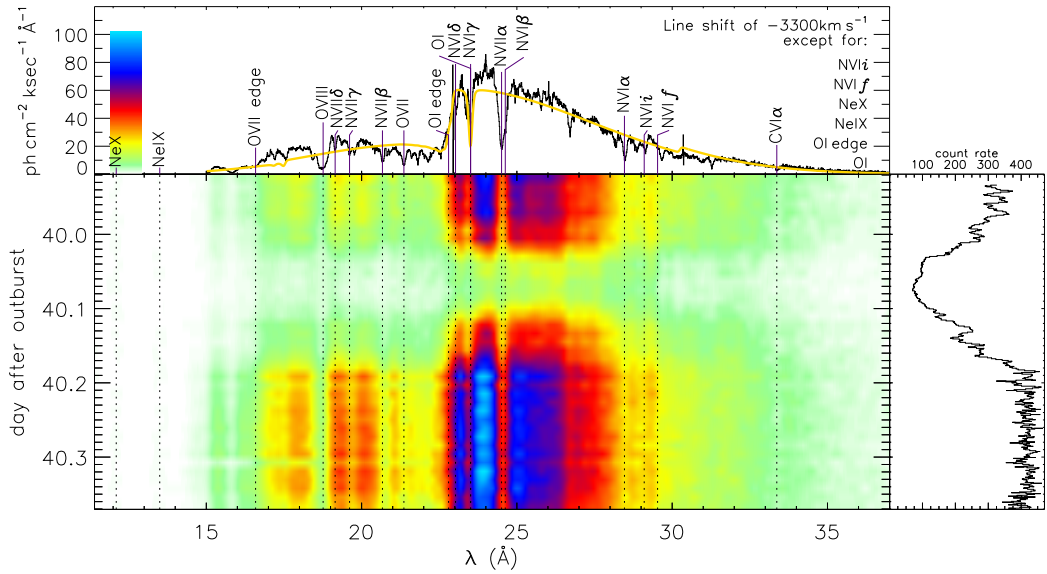}}
\caption{\label{f:v2491cyg_smap}Spectral evolution map following the
same concept as Fig.~\ref{f:rsoph_swmap}, illustrating spectral
changes along the variability in the XMM-Newton observation of
V2491\,Cyg on day 39.9. In the top panel, the total RGS spectrum
is shown with a blackbody fit to the continuum component.
}
\end{figure}

The classical nova KT\,Eri was followed with four Chandra
observations. The nova was extremely bright \citep{kteri_sss}
that extremely short exposures of only a few $10^3$\,seconds
were needed to already obtain extremely well exposed spectra
\citep{ATel2418}. Again, these grating spectra are extremely
complex.\\

\section{Conclusions}
\label{s:concl}

While results from individual observations have already been
discussed in the previous sections, I focus here on more general
conclusions that can be drawn from the entirety of all observations.

\subsection{Pre- and Post-SSS phase}

Originally, novae were only expected to be X-ray sources
during their SSS phase, when the photospheric radius had
receded into the nuclear burning regions near the surface
of the WD. However, already the earliest X-ray observations
of novae have revealed pre-SSS X-ray emission, e.g., in
V838\,Her \citep{obrien94} or V1974\,Cyg \citep{krautt96}.
Plasma models to the low-resolution data indicate that
the emission arises from a collisional plasma and
is suspected to originate in shock fronts within the ejecta
\citep[e.g.,][]{brecher77}. An ``interacting
winds'' model has been developed by
\cite{obrien94} and was refined by \cite{lloyd95}
for the specific case of V838\,Her. They argued
that the shock-heating must take place as a result
of the interaction of different components
{\em within} the ejecta.\\

The much increased sample of X-ray observed novae with
the new-generation X-ray observatories gives an idea
of the frequency of early hard emissions. Out of
27 novae that have been observed with Swift within
the first 100 days after outburst with at least two
observations, 11 were detected but without an SSS component
in the first observation and 17 in the second observation.
Thus about half of all novae display pre-SSS X-ray emission.\\

The interacting winds model predicts inhomogeneities
within the ejecta, and these may have left their footprints
also in the later SSS grating spectra that contain complex
absorption line profiles indicating different velocity
components \citep[see, e.g.,][]{ness_v2491}.
It needs to be kept in mind that the early spectra of
RS\,Oph have a different origin with shocks between
the ejecta and the stellar wind of the companion that is
less dense in main sequence companions. This produces much
more X-ray emission than intra-ejecta shocks.
Owing to their intrinsic faintess, intra-ejecta shock
spectra have never been observed with an X-ray grating.\\

Post-SSS grating spectra were taken of V382\,Vel and of
RS\,Oph, both yielding faint emission line spectra.
While V382\,Vel contains multiple velocity components,
the spectral lines in RS\,Oph are single. Both spectra
can be modeled by collisional plasma models, which is
not necessarily trivial with grating spectra.
The post-SSS emission phase likely originates
in radiatively cooling nebular ejecta
\citep[e.g.,][]{ness_vel} which can be highly non-solar
in composition.

\subsection{SSS phase}
\label{s:grsssresults}

The new observations during the SSS phase have
revealed a number of unexpected complications
such as high-amplitude variations during the early
SSS phase (see Sect.~\ref{s:swmonitoring}),
extremely complex high-resolution spectra,
and a high degree of diversity.\\

The expected form of an SSS spectrum was an
atmospheric absorption line spectrum. However,
the first ever nova that was observed during
the SSS phase with a grating, V1494\,Aql, displayed
a completely different spectrum (see right panel
of Fig.~\ref{f:v1494}), and spectral modelling
was so far unsuccessful \citep{rohrbach09}. The second
observed nova SSS spectrum, of V4743\,Sgr, complied better
with expectations (see Fig.~\ref{f:v4743maps}), but the
light curve was violently variable with the mysterious
decline that only years later can be seen in the context
of the early variability phase. The high degree of
diversity was unexpected, and only with more observations,
some similarities were found, for example a remarkable
similarity between the SSS spectra of RS\,Oph and
V2491\,Cyg \citep{ness_v2491}. Spectral modelling of SSS
spectra still leaves too many doubts to lead to conclusive
results. Not a single approach has so far led to satisfactory
reproduction of the observed spectra in a statistical sense.
Attempts were made with:
\begin{enumerate}
 \item The non-LTE WD atmosphere model TMAP \citep{rauch10}
 \item The expanding radiative transfer model based on PHOENIX \citep{vanRossum2012}
 \item A multicomponent SPEX model \citep{pinto12}.
\end{enumerate}
TMAP is currently the best tested approach and is available to the
scientific community, however, the expansion of the absorbing plasma,
proven by the observed line blue shifts, invalidates the assumptions
of hydrostatic equilibrium and plane parallel geometry. An approach to
compute radiative transport in a co-moving frame has been
implemented in the PHOENIX code, and \cite{vanrossumness09}
have shown that hydrostatic and hydrodynamic models lead
to different spectra, and adjustments to observed spectra
lead to different key parameters. The conclusions drawn from
TMAP modelling are therefore to be treated with care. The PHOENIX
code may be more promising as demonstrated by \cite{vanRossum2012}.
Although currently only models with solar abundances are available,
already good agreement between models and data has been found.
These models will be publicly available and could be used to
parameterise individual grating spectra and then using these
models to interpolate along the low-resolution Swift spectra
to get the full evolution of the atmosphere parameters at the
time density of the Swift monitoring. The approach by
\cite{pinto12} of combining emission- and absorption components
in a self-consistent way may be a promising alternative, but one
has to be careful to use the large number of degrees of freedom
wisely.\\

The studies of U\,Sco have given more clues, as the eclipsing
system displayed an SSS spectrum that resembles the also
eclipsing proto-typical SSS Cal\,87 \citep{ebisawa10}. There is
reason to believe that high-inclination systems contain stronger
emission line components with less-pronounced absorption lines
than low-inclination systems. In Fig.~\ref{f:speccat}, 14 grating
SSS spectra are shown in comparison. The light blue lines guide
the eye that all sources contain photospheric continuum emission
that has the shape of a blackbody. More detailed discussion
will be presented in an upcoming paper Ness et al. in preparation.
An inclination-angle dependence indicates asymmetric ejecta
and possibly the presence of renewed accretion. If we are
dealing with clumpy accretion, the early variability phase
could be explained as temporary ocultations of the central
source that cease when a flatter accretion disk has formed.

\begin{figure}[!ht]
\resizebox{\hsize}{!}{\includegraphics{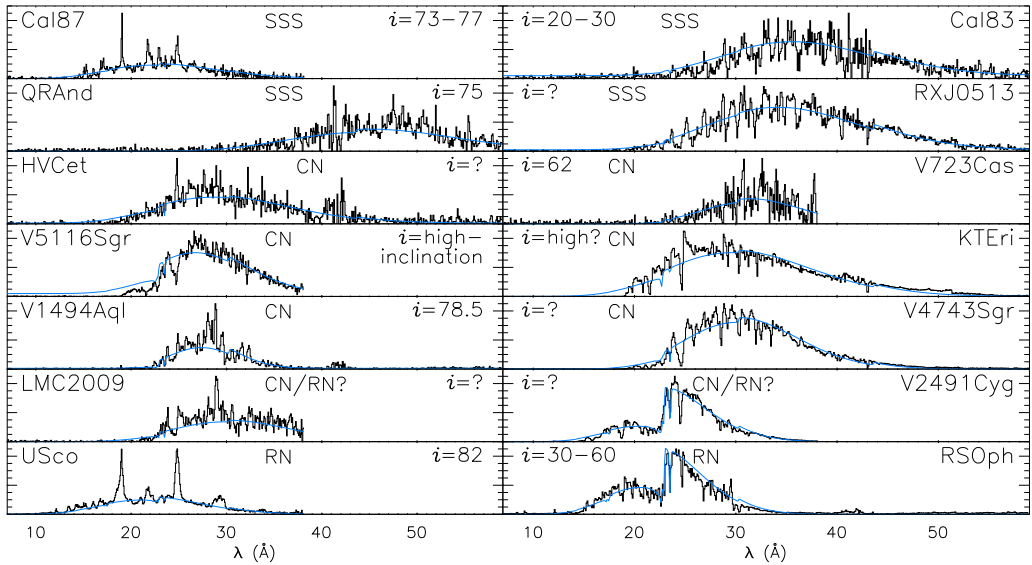}}
\caption{\label{f:speccat}High-resolution X-ray spectra in
arbitrary flux units of supersoft sources that contain emission
lines without absorption features (left column) and those that
consist of continuum and absorption lines (right column).
Inclination angles $i$ from the literature are given if known.
The blue thin lines are blackbody curves,
indicating the presence of photospheric emission in all cases.
The lables SSS, CN, and RN denote persistent supersoft
sources, Classical Novae, and Recurrent Novae, respectively.
More details will be presented in an upcoming paper by
Ness et al.
}
\end{figure}


\bibliographystyle{mn2e}
\bibliography{cn,astron,jn,rsoph,usco,vel}


\appendix

\label{lastpage}
\end{document}